\documentclass[aps,prb,preprint]{revtex4-1}
\usepackage[utf8]{inputenc}
\usepackage{amsmath}
\usepackage{bm}
\usepackage{graphicx}
\usepackage{hyperref}
\usepackage{dsfont}
\usepackage[capitalize]{cleveref}
\usepackage[acronyms]{glossaries}
\usepackage{pgfplots}

\graphicspath{{figures/}}

\usetikzlibrary{arrows,arrows.meta}
\tikzset{>=latex, every picture/.style=thick}

\makeatletter
\@ifclasswith{revtex4-1}{reprint}
{
  \newcommand{\figscale}{0.92}}
{
  \newcommand{\figscale}{0.85}}
\makeatother

\hypersetup{
  colorlinks,
  linkcolor=blue,
  urlcolor=blue,
  citecolor=blue,
  pdftitle={Density-matrix functional theory of the attractive Hubbard
    model: Statistical analogy of pairing correlations},
  pdfauthor={T. S. Müller and G. M. Pastor}
}

\renewcommand{\cite}[1]{[\onlinecite{#1}]}

\newcommand{\lr}[1]{\ensuremath{\langle #1 \rangle}}       
\newcommand{\ct}[2][c]{\ensuremath{\hat{#1}_{#2}^\dagger}}   
\newcommand{\ca}[2][c]{\ensuremath{\hat{#1}_{#2}}}         
\newcommand{\bra}[1]{\ensuremath{\langle#1|}}         
\newcommand{\ket}[1]{\ensuremath{|#1\rangle}}         

\newcommand{\HF}{\ensuremath{\mathrm{HF}}}

\newcommand{\oo}{\ensuremath{\infty}}                 
\newcommand{\veps}{\ensuremath{\varepsilon}}          
\newcommand{\up}{\ensuremath{\uparrow}}               
\newcommand{\dw}{\ensuremath{\downarrow}}             
\newcommand{\Tr}{\ensuremath{\mathrm{Tr}}}            
\newcommand{\eff}{\ensuremath{\mathrm{eff}}}          
\newcommand{\ksig}{\ensuremath{\bm{k}\sigma}}         
\newcommand{\IFE}{\ensuremath{\mathrm{IFE}}}          
\newcommand{\ex}{\ensuremath{\mathrm{ex}}}            

\glsdisablehyper{}

\newacronym{dft}{DFT}{density-functional theory}
\newacronym{ldft}{LDFT}{lattice density-functional theory}
\newacronym{rdmft}{RDMFT}{reduced density matrix functional theory}
\newacronym{1rdm}{1RDM}{one-particle reduced density matrix}
\newglossaryentry{xc}{name={XC}, description={exchange-correlation},
  first={exchange and correlation~(XC)}, user1={exchange-correlation~(XC)},
  type=\acronymtype}
\newacronym{ftdft}{FT-DFT}{finite-temperature density functional theory}
\newacronym{dos}{DOS}{density of states}
\newacronym{hf}{HF}{Hartree-Fock}
\newglossaryentry{tf}{name={TF}, description={Thomas-Fermi},
  first={Thomas and Fermi~(TF)}, type=\acronymtype}
\newacronym{df}{DF}{density functional}
\newglossaryentry{hk}{name={HK}, description={Hohenberg-Kohn},
  first={Hohenberg and Kohn~(HK)}, user1={Hohenberg-Kohn~(HK)}, type=\acronymtype}
\newglossaryentry{ks}{name={KS}, description={Kohn-Sham},
  first={Kohn and Sham~(KS)}, user1={Kohn-Sham~(KS)}, type=\acronymtype}
\newacronym{lda}{LDA}{local density approximation}
\newacronym{lsda}{LSDA}{local spin-density approximation}
\newacronym{gga}{GGA}{generalized gradient appro{x}imation}
\newacronym{ll}{LL}{Levy-Lieb}
\newacronym{dmft}{DMFT}{dynamical mean-field theory}
\newacronym[firstplural=single-particle density matrices (SPDMs)]{spdm}{SPDM}%
           {single-particle density matrix}
\newacronym{gsspdm}{gs-SPDM}{ground-state single-particle density matrix} 
\newacronym{eqspdm}{eq-SPDM}{equilibrium single-particle density matrix}
\newacronym{ftldft}{FT-LDFT}{finite-temperature lattice density functional theory}
\newacronym{ife}{IFE}{independent-fermion entropy}
\newacronym{bz}{BZ}{first Brillouin zone}
\newacronym{nn}{NN}{nearest neighbor}
\newacronym{fm}{FM}{ferromagnetic}
\newacronym{pm}{PM}{paramagnetic}
\newacronym{afm}{AFM}{antiferromagnetic}
\newacronym{qmc}{QMC}{quantum Monte Carlo}
\newacronym{bcs}{BCS}{Bardeen, Cooper, and Schrieffer}
\newacronym{nlce}{NLCE}{numerical linked-cluster expansion}
\newacronym{cs}{CS}{constrained search}
\newacronym{cdw}{CDW}{charge-density wave}

\begin{document}

\title{Density-matrix functional theory of the attractive Hubbard model:
  Statistical analogy of pairing correlations}

\author{T. S. Müller}
\author{G. M. Pastor}
\affiliation{%
  Institut für Theoretische Physik, Universität Kassel,
  Heinrich-Plett-Strasse 40, 34132 Kassel, Germany}
\date{\today}
\begin{abstract}
  The ground-state properties of the Hubbard model with attractive
  local pairing interactions are investigated in the framework of
  \acrlong{ldft}. A remarkable correlation is revealed between the
  interaction-energy functional $W[\bm{\eta}]$ corresponding to the
  Bloch-state occupation-number distribution $\eta_{\ksig}$ and the
  entropy $S[\bm{\eta}]$ of a system of non-interacting fermions
  having the same $\eta_{\ksig}$. The relation between $W[\bm{\eta}]$
  and $S[\bm{\eta}]$ is shown to be approximately linear for a wide
  range of ground-state representable occupation-number distributions
  $\eta_{\ksig}$. Taking advantage of this statistical analogy, a
  simple explicit ansatz for $W[\bm{\eta}]$ of the attractive Hubbard
  model is proposed, which can be applied to arbitrary periodic
  systems. The accuracy of this approximation is demonstrated by
  calculating the main ground-state properties of the model on several
  1D and 2D bipartite and non-bipartite lattices and by comparing the
  results with exact diagonalizations.
\end{abstract}

\maketitle
\newpage

\glsresetall{}

\section{Introduction}%
\label{sec:intro}

The study of strongly correlated phenomena in quantum many-body
systems is one of the central challenges in condensed-matter
physics. Over the past decades, the field has experienced a remarkable
expansion which has been fueled by the discovery of new materials, as
well as by the progress in experimental characterization tools and
theoretical methodologies. One of the most significant developments in
this field has been the discovery of high temperature superconductors
and the novel electronic pairing mechanism associated with
them~\cite{Dagotto_jul1994, Bednorz_jun1986}. Indeed, in these
materials the origin of superconductivity is profoundly different from
the phonon-mediated coupling at the center of the \gls{bcs} theory of
conventional superconductors~\cite{Bardeen_1957}.  As a result, the
study of alternative descriptions of pairing interactions in solids
have gained increasing attention~\cite{Bickers_aug1987,
  Baskaran_jan1988, Kotliar_sep1988, Schrieffer_jun1989,
  Kampf_apr1990, Anderson_1991, Dagotto_jul1994, Anderson_jul1998}.

The attractive single-band Hubbard model~\cite{Hubbard_nov1963,
  Kanamori_jan1963, Gutzwiller_mar1963} with on-site
interactions~$U<0$ provides a simple, albeit oversimplified way of
introducing a pairing mechanism among fermions and of exploring its
consequences on the many-body properties~\cite{Hirsch_jul1985,
  Hirsch_jun1986, Scalettar_mar1989, Paiva_may2004,
  Saubanere_sep2014}. Originally proposed for describing local
repulsive Coulomb interactions in the context of itinerant narrow-band
magnetism~\cite{Hubbard_nov1963, Kanamori_jan1963,
  Gutzwiller_mar1963}, this model has played, together with other
lattice models~\cite{Stoner_apr1938, Anderson_oct1961, Kondo_jul1964},
a most significant role in sharping our understanding of strongly
correlated phenomena.  Therefore, it is most interesting to explore
its physical properties when pairing is favored. From this perspective
it is important to recall that the physics described by the attractive
Hubbard model is intrinsically different from the pairing mechanism of
the \gls{bcs}~theory. The effective interactions in the
\gls{bcs}~theory have an off-diagonal character since they are
mediated by electron-phonon scattering. Moreover, the narrow energy
and momentum dispersion caused by the interaction with phonons results
in a large spatial extension of the Cooper pairs. In contrast, the
pairing interactions are strictly local in the Hubbard model, since
they only affect fermions occupying the same lattice site. Rather than
a limitation, this strong complementarity is one of the reasons which
renders the model particularly appealing from a theoretical
perspective~\cite{Micnas_jan1990}.

Local attractive interactions have been the subject of multiple
theoretical studies by using a variety of
methodologies~\cite{Hirsch_jul1985, Hirsch_jun1986, Scalettar_mar1989,
  Micnas_jan1990, Marsiglio_jan1997, Tanaka_aug1999, Paiva_may2004,
  Salwen_aug2004, Hu_jul2010, Saubanere_sep2014, Schilling_dec2018,
  Schilling_jan2019, Schmidt_jun2019, Benavides-Riveros_may2020,
  Schmidt_sep2021}. In early works the formation and stability
of \glspl{cdw} and superconducting states has been
quantified~\cite{Hirsch_jul1985, Hirsch_jun1986}. In addition, the
phase diagram of the two-dimensional (2D) negative-$U$ Hubbard model
has been determined~\cite{Scalettar_mar1989, Paiva_may2004}. Other
investigations have addressed the accuracy of the mean-field
\gls{bcs}~approximation by comparing it with the exact Bethe-ansatz
solution of the one dimensional (1D) Hubbard model as well as with
accurate numerical calculations in two
dimensions~\cite{Marsiglio_jan1997, Tanaka_aug1999, Salwen_aug2004,
  Hu_jul2010}. More recently, the model has been studied in the
framework of \gls{ldft}~\cite{Saubanere_sep2014} by applying the
concepts of first-principles
\gls{dft}~\cite{Hohenberg_nov1964,Parr_1995} to many-body lattice
Hamiltonians. An explicit semilocal approximation to the interaction
energy $W$ has been proposed as a functional of the
\gls{spdm}~$\bm{\gamma}$. In this way, the ground-state kinetic,
Coulomb and total energies, the charge distribution and
nearest-neighbor bond order as well as the pairing energy have been
determined for various 1D, 2D and 3D
lattices~\cite{Saubanere_sep2014}. Moreover, even-odd and super-even
oscillations of the pairing energy have been observed as a function of
the band filling in agreement with previous
studies~\cite{Tanaka_aug1999}.

The purpose of this paper is to investigate the properties of the
attractive Hubbard model in the framework of \gls{ldft}. As a
complement to the previous studies~\cite{Saubanere_sep2014}, which are
based on the exact solution of the two-site problem and the scaling
properties of $W[\bm{\gamma}]$, we adopt here a delocalized
$\bm{k}$-space perspective. Thus, the translational symmetry of the
lattice allows us to regard the interaction energy $W$ as a functional
of the occupation-number distribution~$\eta_{\ksig}$ of the Bloch
states having wave vector~$\bm{k}$. As in any density-functional
approach, the accuracy of the outcome relies on the quality of the
considered approximation to the functional $W[\bm{\eta}]$, where
$\bm{\eta}$ denotes the vector with components
$\eta_{\ksig}$. Therefore, our first goal is to propose an appropriate
explicit ansatz for $W[\bm{\eta}]$, which in our case is based on a
statistical interpretation of pairing correlation effects. An
analogous approach has been recently proven to be quite successful for
repulsive interactions~\cite{Muller_jul2018}. Once the functional is
introduced, and its capacity to correctly describe attractive
interactions is examined, we proceed to a number of applications of
\gls{ldft} to one- and two-dimensional negative-$U$ Hubbard models in
order to assess the quantitative accuracy of the method and to discuss
its goals and limitations.

The remainder of the paper is organized as follows. In \cref{sec:theo}
the main concepts of \gls{ldft} are recalled. This includes discussing
the central role played by the \gls{spdm}~$\bm{\gamma}$ and the
Bloch-state occupation-number distribution $\eta_{\ksig}$, as well as
introducing the variational principle from which the ground-state
properties are derived. In \cref{sec:analogy} a remarkable correlation
is revealed between the interaction energy~$W$ of the attractive
Hubbard model and the entropy~$S$ of a system of noninteracting
fermions, both regarded as functionals of $\eta_{\ksig}$. Based on
this statistical analogy, we formulate the so-called
\gls{ife}~approximation to the interaction-energy functional
$W[\bm{\eta}]$. Applications of this ansatz to the attractive Hubbard
model on one- and two-dimensional lattices are presented and discussed
in \cref{sec:results}.  Comparison with exact numerical
calculations underscores the accuracy and predictive power of the
method. Finally, the paper is closed in \cref{sec:summary} with a
summary of our conclusions.

\section{Theoretical Background}%
\label{sec:theo}

Consider a many-body lattice Hamiltonian consisting of a single-particle
kinetic energy operator $\hat{T}$ and a two-particle interaction operator
$\hat{W}$:
\begin{equation}
  \label{eq:1}
  \hat{H} = \hat{T} + \hat{W}
  = \sum_{ij\sigma} t_{ij\sigma} \, \ct{i\sigma} \ca{j\sigma}
  + \frac{1}{2}\sum_{\substack{ijkl\\\sigma\sigma'}}
  W_{ijkl}^{\sigma\sigma'} \, \ct{i\sigma}\ct{j\sigma'}
  \ca{l\sigma'}\ca{k\sigma} \,,
\end{equation}
where $\ct{i\sigma}$ ($\ca{i\sigma}$) creates (annihilates) an
electron with spin $\sigma$ in the orbital $\phi_i(\bm{r})$. In the
case of single-band models as the Hubbard model, the index $i$
corresponds simply to the lattice site, while in multiband models
(e.g., $d$-band models) it also labels the different local orbitals
which are taken into account. The parameters in $\hat{H}$ are the
single-particle energy levels $t_{ii\sigma}$ and hopping integrals
$t_{ij\sigma}$ with $i\neq j$, which define $\hat{T}$, and the
interaction integrals $W_{ijkl}^{\sigma\sigma'}$, which define
$\hat{W}$. Once the orbitals and the basic characteristics of the
model interactions $W_{ijkl}^{\sigma\sigma'}$ are adopted, the problem
is entirely set by the the single-particle matrix elements
$t_{ij\sigma}$. In particular they define the dimensionality and
topology of the lattice, the range of the single-particle
hybridizations and the relative importance between kinetic and
interaction energies which conditions the nature of the correlations.

Since the hoppings $t_{ij\sigma}$ enter the Hamiltonian in a bilinear
form together with the operators $\ct{i\sigma} \ca{j\sigma}$, it is
possible to replace the wave-function $\ket{\Psi}$ or the mixed-state
density-matrix $\hat{\rho}$ by the single-particle density matrix
$\bm{\gamma}$, whose elements are
$\gamma_{ij\sigma} = \langle \ct{i\sigma} \ca{j\sigma}\rangle$, as the central
unknown of the many-body problem~\cite{Lopez-Sandoval_jan2000,Tows2014_tca}.
The situation is analogous to the one found in the first-principles theory
of the inhomogeneous electron gas, where the external potential $v(\bm{r})$
defines the problem and therefore the electronic density $\rho(\bm{r})$
becomes the central variable of density-functional
theory~\cite{Hohenberg_nov1964}. A proof of the Hohenberg-Kohn theorem
for lattice models may be found in Ref.~\cite{Tows_jun2011}.

Starting from the ground-state variational principle and following
Levy and Lieb's two-step minimization~\cite{Levy_sep1982,
  Lieb_sep1983}, it is easy to show that the ground state energy of
the many-body Hamiltonian $\hat{H}$ is given by
\begin{equation}
  \label{eq:2}
  E_0 = \min_{\bm{\gamma}\in\bm{\Gamma}_N} \bigg\{ \sum_{ij\sigma}
    t_{ij\sigma} \, \gamma_{ij\sigma} + W[\bm{\gamma}] \bigg\} \,,
\end{equation}
where the minimization runs over the set $\bm{\Gamma}_N$ of all
physical single-particle density matrices, i.e., over all
$\bm{\gamma}$ which can be obtained from an $N$-particle wave function
$\ket{\Psi}$ or a mixed state $\hat{\rho}$. The interaction energy
$W[\bm{\gamma}]$ is a universal functional of $\bm{\gamma}$, i.e., a
functional which is independent of the hopping integrals defining the
specific problem under consideration. Of course, $W[\bm{\gamma}]$
depends on the form and nature of the interactions
$W_{ijkl}^{\sigma\sigma'}$ and on the number of particles $N$ which is
given by $\bm{\gamma}$ itself ($N = \sum_{i\sigma} \gamma_{ii\sigma}$).
Physically, $W[\bm{\gamma}]$ represents the minimum value that the
interaction energy of the many-body system can take when the
single-particle density matrix is equal to $\bm{\gamma}$. This can be clearly
seen from its formal expression~\cite{Valone_aug1980,Valone_nov1980,%
Lopez-Sandoval_jan2000,Tows_jun2011}
\begin{equation}
  \label{eq:3}
  W[\bm{\gamma}] =
  \min_{\hat{\rho}\to\bm{\gamma}} \Tr\big\{\hat{\rho}\,\hat{W}\big\} \,,
\end{equation}
where the minimization runs over all $N$-particle mixed states
$\hat{\rho}$ yielding
$\langle \ct{i\sigma} \ca{j\sigma}\rangle = \gamma_{ij\sigma}$ for all
$ij\sigma$. While the physical interpretation of $W[\bm{\gamma}]$ is
transparent and sound, particularly in combination with \cref{eq:2},
finding its exact functional dependence is far from simple, since
$W[\bm{\gamma}]$ conceals, among other information, the interaction
energy in the ground state of $\hat{H}$ for all possible band fillings
and hopping integrals $t_{ij\sigma}$~\footnote{The domain of
  definition of $W[\bm{\gamma}]$ comprises not only density matrices
  which can be derived from some ground-state of the Hubbard model
  (i.e., the so-called ground-state representable $\bm{\gamma}$) but
  the much broader set of ensemble-representable density matrices,
  which can be derived from an arbitrary mixed state $\hat{\rho}$. The
  necessary and sufficient condition for a density matrix to be
  ensemble representable is simply that all its eigenvalues
  $\eta_{\bm{k}\sigma}$ are comprised between zero and one.}. As in
the theory of the inhomogeneous electron gas, the challenge in
\gls{ldft} is to find accurate explicit approximations to
$W[\bm{\gamma}]$. Once such an approximation is available, \cref{eq:2}
implies that the ground-state energy and density matrix can be
obtained by minimizing the energy functional
\begin{equation}
  \label{eq:4}
  E[\bm{\gamma}] = T[\bm{\gamma}] + W[\bm{\gamma}]
        = \sum_{ij\sigma} t_{ij\sigma} \, \gamma_{ij\sigma} + W[\bm{\gamma}]
\end{equation}
with respect to $\bm{\gamma}$, which can thus be regarded as the central
variable of the many-body
problem~\cite{Lopez-Sandoval_jan2000,Tows2014_tca}.

In periodic solids it is meaningful to take advantage of translational
symmetry and restrict the domain of minimization to translational
invariant density matrices. In this case
\begin{equation}
  \label{eq:5}
  \gamma_{ij\sigma} = \sum_{\bm{k}} u_{i\ksig} \, \eta_{\ksig}
  \, u_{j\ksig}^{*} \, ,
\end{equation}
where the eigenvectors $\bm{u}_{\ksig}$ of $\bm{\gamma}$ are Bloch
states and the eigenvalues $\eta_{\ksig}$ represent the corresponding
occupation numbers. Note that in single-band models $\bm{k}$ stands
simply for the Bloch wave-vector in the first Brillouin zone, whereas
in multiband models it also implicitly includes the band index. In the
former case (one orbital per unit cell) the eigenvectors are entirely
defined by translational symmetry and can thus be classified by the
Bloch wave-vector $\bm{k}$. This represents a significant
simplification which applies in particular to the Hubbard model. It
means that the kinetic and interaction energy can be regarded as
functionals $T[\bm{\eta}]$ and $W[\bm{\eta}]$ of the occupation number
distribution $\eta_{\ksig}$ alone. Only in the single-band case the
shape of the ground-state density-matrix eigenvectors can be assumed
to be independent of the parameters defining the physical model (e.g.,
band filling, interaction strength, etc.).

\section{Statistical analogy}%
\label{sec:analogy}

In order to develop an explicit practical approximation to the
interaction-energy functional $W[\bm{\eta}]$ we focus on the
attractive single-band Hubbard model
\begin{equation}
  \label{eq:6}
  \hat{H} = \hat{T} + \hat{W}
  = \sum_{ij\sigma} t_{ij} \, \ct{i\sigma} \ca{j\sigma}
  + U \sum_i \hat{n}_{i\up} \hat{n}_{i\dw} \,,
\end{equation}
where $t_{ij}$ stands for the hopping integral between sites~$i$
and~$j$, $U \le 0$ is the local interaction strength and
$\hat{n}_{i\sigma} = \ct{i\sigma} \ca{i\sigma}$ is the number operator
for spins $\sigma$ at site~$i$. Consequently, the interaction-energy
functional according to \cref{eq:3} is given by
\begin{equation}
  \label{eq:7}
  W[\bm{\gamma}] =
  \min_{\hat{\rho}\to\bm{\gamma}} \Big\{ U \sum_i \lr{\hat{n}_{i\up}
    \hat{n}_{i\dw}} \Big\}
  = U \, \max_{\hat{\rho}\to\bm{\gamma}} \Big\{
  \sum_i \lr{\hat{n}_{i\up}\hat{n}_{i\dw}}\Big\}\,,
\end{equation}
where we have used that $U\le 0$. Two important limiting cases, for
which Levy's constrained minimization~\eqref{eq:7} can be exactly
solved, are worth considering from the start. The first one is the
scalar spin-density matrix
$\gamma_{ij\sigma} = \delta_{ij} n_{\sigma}$, where the density
$n_{\sigma} = N_{\sigma}/N_a$ and total number $N_{\sigma}$ of
spin-$\sigma$ fermions are the same for both spins and $N_a$ denotes
the number of atoms. The corresponding occupation numbers
$\eta_{\ksig} = n_{\sigma}$ are then independent of~$\bm{k}$ and
$\sigma$. In this case the interaction energy $W[\bm{\eta}]$ assumes
its minimal value $W_{\oo} = UD_{\oo}$ by adopting a fully localized
state with the maximum number of pairs or double occupations
$D_{\oo} = N_{\up}=N_{\dw}$. The second important limit is defined by
the idempotent density matrices $\bm{\gamma} = \bm{\gamma}^2$, which
have $\eta_{\ksig}=0$ or $1$ for all $\ksig$
($\sum_{\ksig} \eta_{\ksig} = N_{\sigma}$). The many-particle states
by which idempotent \glspl{spdm} can be represented are the Slater
determinants made of the Bloch states having $\eta_{\ksig}=1$. The
corresponding interaction-energy $W$ is then given by the
\gls{hf}~energy
\begin{equation}
  \label{eq:8}
  W_{\HF} = U \sum_i \gamma_{ii\up} \gamma_{ii\dw} = U N_a \, n_{\up}
  n_{\dw} \,,
\end{equation}
where we have used that $\gamma_{ii\sigma}=n_{\sigma}$ for all~$i$. In
this context it is useful to draw the following statistical analogy,
by regarding $\eta_{\ksig}$ as the average occupation numbers of
$N_{\up}$ and $N_{\dw}$ fermions which are statistically distributed
over all the Slater determinants that can be constructed with the
single-particle Bloch states~\cite{Collins1993,Muller_jul2018}.
In the fully localized state, where
pairing is maximal, $\eta_{\ksig} = n_{\sigma}$ is uniform and
therefore all Bloch states are equally probable. The occupation-number
distribution $\eta_{\ksig}$ conceals no a priori information on which
Bloch states are occupied and which are empty. The information content
of this distribution is zero and the entropy is maximal. On the other
extreme, for an idempotent $\bm{\gamma}$, $\eta_{\ksig}$ is equal to 0
or 1, which means that only one Slater determinant is possible. Our
knowledge on the distribution of the system across its Slater determinants
is complete and the entropy vanishes~\cite{Collins1993,Muller_jul2018}.
This suggests that the entropy
\begin{equation}
  \label{eq:9}
  S[\bm{\eta}] = -\sum_{\ksig} \big[
  \eta_{\ksig} \ln(\eta_{\ksig}) + (1 - \eta_{\ksig})
  \ln(1-\eta_{\ksig})
  \big]
\end{equation}
of a system of independent fermions having the occupation-number
distribution $\eta_{\ksig}$ can be regarded as a measure of the
ability of the system to minimize the interaction energy under the
constraint of the given $\eta_{\ksig}$ [see \cref{eq:7}].  Notice that
the \acrlong{ife}~(\acrshort{ife}) assumes its extremes for the same
$\eta_{\ksig}$ as the interaction energy~$W$: If the fermions are
uncorrelated ($\eta_{\ksig}=0$ or 1) $W=W_{\HF}$ is maximal and $S=0$
is minimal. On the other hand, if the fermions are fully paired and
localized ($\eta_{\ksig} = n_{\sigma}$ for all~$\bm{k}$)
$W=W_{\oo}=U N_{\up} = U N_{\dw}$ is minimal and
\begin{equation}
  \label{eq:10}
  S = S_{\oo} =
  -N_a \sum_{\sigma} \big[n_{\sigma} \ln(n_{\sigma}) +
  (1-n_{\sigma}) \ln(1-n_{\sigma})\big]
\end{equation}
is maximal.  These correlations suggest that $S[\bm{\eta}]$, which is
itself a functional of the occupation-number distribution $\bm{\eta}$,
can be used for deriving an effective approximation of $W[\bm{\eta}]$
~\cite{Collins1993,Muller_jul2018}%
\footnote{
In Ref.~\cite{Collins1993} the importance of the growing statistical
uncertainty in the occupation-number distribution resulting from increasing
correlations has been clearly identified. The arguments presented here
in order to derive Eq.~(\ref{eq:9}) in the weakly correlated limit are
very similar to those formulated in Collins' work. However, Eq.~(\ref{eq:9})
is profoundly different physically from the information
entropy $S = -\sum_k \eta_k \ln(\eta_k)$ considered in Ref.~\cite{Collins1993}.
Eq.~(\ref{eq:9}) represents the entropy of a many-body system of
non-interacting fermions having the same occupation-number distribution
$\eta_{k\sigma}$ as the interacting system under consideration. Thus,
the fermionic character of the particles is correctly taken into account
and the fundamental electron-hole symmetry of $W[\bm{\eta}]$ is respected,
which does not hold for Collins' conjecture.
}.

In order to quantify the actual correlation between $S[\bm{\eta}]$ and
$W[\bm{\eta}]$ we have performed exact numerical Lanczos
diagonalizations of the ground state of the attractive Hubbard model
for a number of different lattice structures and band fillings. A wide
variety of occupation-number distributions $\bm{\eta}$ has
been explored by varying the hopping integrals from $t_{ij}=0$ to
$|t_{ij}|\gg |U|$ so as to scan $S[\bm{\eta}]$ in its complete range
($0\le S\le S_{\oo}$) and by considering different ratios and relative signs
between first and second NN hoppings.
\begin{figure}[ptb]
  \centering
  \includegraphics[scale=\figscale]{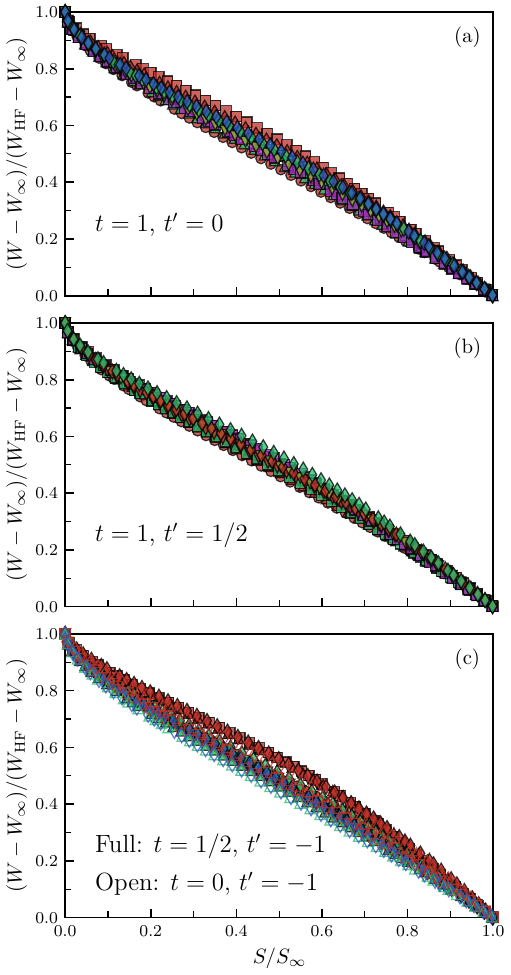}
  \caption{
    Correlation between the interaction energy~$W$ [\cref{eq:7}] and
    the \acrlong{ife} $S$ [\cref{eq:9}] in the ground state of the attractive
    Hubbard model for different structures, sizes and band fillings of
    vanishing spin polarization ($N_\uparrow = N_\downarrow$). In (a) only
    first \gls{nn}-hoppings $t_{ij} = -t < 0$ are taken into account:
    1D rings having a number of sites
    $N_a = 6$ (circles), $10$ (upright triangles)
    and $14$ (squares), as well as 2D square lattices having
    $N_a = 2 \times 4$ (downright triangles) and $N_a = 3 \times 4$ (diamonds).
    The considered numbers of spin-$\sigma$ fermions are
    $N_\sigma = 3$ (red), $4$ (yellow), $5$ (green),
    $6$ (blue) and $7$ (magenta).
    In (b) second \gls{nn}-hoppings $t_{ij} = -t' = -t/2 < 0$ are included
    for the same structures and band fillings as in (a).
    In (c) results are shown for 1D rings with
    $N_a = 6$ (circles), $7$ (squares), $8$ (downright triangles),
    $10$ (upright triangles) and $12$ sites (diamonds) having
    strongly competing hopping integrals:
    first \gls{nn}-hoppings $t_{ij} = -t < 0$ and second
    \gls{nn}-hoppings $t_{ij} = -t'= 2t > 0$ (full symbols) as well as 
    first \gls{nn}-hoppings $t_{ij} = 0$ and second
    \gls{nn}-hoppings $t_{ij} =  -t'> 0$ (open symbols) for
    $N_\sigma = 2$ (red), $4$ (green) and $6$ (blue).}%
  \label{fig:1}
\end{figure}
The results shown in \cref{fig:1} reveal a remarkable nearly
one-to-one correspondence between $W[\bm{\eta}]$ and $S[\bm{\eta}]$ in
all considered situations including bipartite and non-bipartite 1D and
2D lattices, as well as systems having competing first and second NN
hoppings with unusual single-particle dispersion relations.
Moreover, once properly scaled
($W_{\oo} \le W \le W_{\HF}$ and $0\le S \le S_{\oo}$) the relation
becomes approximately independent of the size and dimension of the
system under consideration. This pseudo-universal relation between $W$
and $S$ provides an ideal ground for broadly applicable approximations
to the interaction-energy functional $W[\bm{\eta}]$. On the basis of
the numerical results shown in \cref{fig:1} we propose the linear
ansatz
\begin{equation}
  \label{eq:11}
  W[\bm{\eta}] = W_{\HF} + (W_{\oo}-W_{\HF}) \,
  \frac{S[\bm{\eta}]}{S_{\oo}}
\end{equation}
for the interaction-energy functional of the attractive Hubbard model with
$N_{\up}=N_{\dw}$ fermions. An analogous approximation has already been
proposed for the repulsive Hubbard model at half-band filling
($N_{\up}+N_{\dw}=N_a$)~\cite{Muller_jul2018}. A different proportionality
relation between the correlation energy and the information entropy of the
distribution of occupation numbers $\eta_k$ has been previously proposed
by Collins~\cite{Collins1993} and more recently applied by other
authors~\cite{Flores2016,Wang2021} in atomic and molecular 
calculations~\cite{Note2}.

It is instructive to contrast the interaction-energy functional of the
Hubbard model for attractive interactions $U<0$, which we denote in
the following by $W^-[\bm{\gamma}]$, with the corresponding functional
of the repulsive case, which we denote by
$W^+[\bm{\gamma}]$. Establishing any possible relation between $W^+$
and $W^-$ would be helpful in order to guide the search for accurate
approximations or to verify the validity of new ones. Consider the set
$\Omega(\gamma_{ij\up},\gamma_{ij\dw})$ of all normalized pure states
$\ket{\Psi}$, or for that matter mixed states $\hat{\rho}$, yielding
the density matrix $\bm{\gamma}$, i.e.,
$\bra{\Psi} \ct{i\sigma}\ca{j\sigma}\ket{\Psi}=\gamma_{ij\sigma}$ for
all $\ket{\Psi} \in \Omega(\gamma_{ij\up},\gamma_{ij\dw})$. This set
is not empty since $\bm{\gamma}$ is physical, i.e.,
$N$-representable. One may then identify the states
$\ket{\Psi_{\max}}$ and $\ket{\Psi_{\min}}$ in
$\Omega(\gamma_{ij\up},\gamma_{ij\dw})$ having the largest,
respectively smallest, total number of double occupations $D$. Let
$D_{\max} = \bra{\Psi_{\max}} \sum_i \hat{n}_{i\up} \hat{n}_{i\dw}
\ket{\Psi_{\max}}$ and
$D_{\min} = \bra{\Psi_{\min}} \sum_i \hat{n}_{i\up} \hat{n}_{i\dw}
\ket{\Psi_{\min}}$ be the corresponding maximum and minimum $D$ within
$\Omega$, which are all functionals of $\bm{\gamma}$. There is no
simple relation known to us between $D_{\max}$ and $D_{\min}$ for an
arbitrary $\bm{\gamma}$. However, it is easy to conceive a bijective
mapping of the set $\Omega$ onto an in general different set
$\Omega^h$ in which $\ket{\Psi_{\min}}$ is assigned to
$\ket{\Psi_{\max}}$ and vice versa. This is achieved by performing an
electron-hole transformation on one of the spin components. For each
$\ket{\Psi}\in\Omega$, one obtains its image
$\ket{\Psi^h} \in \Omega^h$ by replacing the fermion creation
operators $\ct{i\sigma}$ by the hole operators
$\ct[h]{i\up}=\ct{i\up}$ and $\ct[h]{i\dw}=\ca{i\dw}$ and by replacing
the vacuum state by the ket in which all down-spin orbitals are
occupied. It follows that the density matrix $\bm{\gamma}^h$ of any
state in $\Omega^h$ is given by
\begin{align}
  \label{eq:12}
  \gamma_{ij\up}^h
  &= \bra{\Psi^h} \ct{i\up} \ca{j\up} \ket{\Psi^h}
  = \bra{\Psi^h} \ct[h]{i\up} \ca[h]{j\up} \ket{\Psi^h}
    = \gamma_{ij\up}\\
  \label{eq:13}
  \gamma_{ij\dw}^h
  &= \bra{\Psi^h} \ct{i\dw} \ca{j\dw} \ket{\Psi^h}
  = \bra{\Psi^h} \ca[h]{i\dw} \ct[h]{j\dw} \ket{\Psi^h}
  = \delta_{ij} - \gamma_{ji\dw} \,.
\end{align}
Therefore,
$\Omega^h \equiv \Omega(\gamma_{ij\up},
\delta_{ij}-\gamma_{ji\dw})$. Moreover, the double-occupation
operators transform as
\begin{equation}
  \label{eq:14}
  \hat{n}_{i\up}\hat{n}_{i\dw} =
  \hat{n}_{i\up}^h (1-\hat{n}_{i\dw}^h) = \hat{n}_{i\up}^h -
  \hat{n}_{i\up}^h\hat{n}_{i\dw}^h \,,
\end{equation}
which implies that the total number
of double occupations in $\ket{\Psi^h}$ is given by
\begin{equation}
  \label{eq:15}
  \begin{aligned}
    D^h &=
    \bra{\Psi^h} \sum_i \hat{n}_{i\up} \hat{n}_{i\dw} \ket{\Psi^h}\\
    &= \bra{\Psi^h} \sum_i \big( \hat{n}_{i\up}^h -
    \hat{n}_{i\up}^h\hat{n}_{i\dw}^h \big) \ket{\Psi^h} =
    N_{\up} - D \,,
  \end{aligned}
\end{equation}
where $D$ is the total number of double occupations in
$\ket{\Psi}$. Since $N_{\up} = \sum_i \gamma_{ii\up}$ is the same for
all $\ket{\Psi}$ and $\ket{\Psi^h}$, the change of sign in $D$ means
that the state $\ket{\Psi_{\min}}$ yielding the minimum number of
double occupations $D_{\min}$ in
$\Omega(\gamma_{ij\up},\gamma_{ij\dw})$ is mapped into the state
$\ket{\Psi_{\max}^h}$ yielding the maximum $D_{\max}^h$ in
$\Omega(\gamma_{ij\up}, \delta_{ij}-\gamma_{ji\dw})$ and vice
versa. One concludes that the interaction energy functionals $W^-$,
corresponding to the attraction $U<0$, and $W^+$, corresponding to the
repulsion $-U>0$, are in general related by
\begin{equation}
  \label{eq:16}
  W^-[\gamma_{ij\up},\delta_{ij}-\gamma_{ji\dw}]
  = U \sum_i \gamma_{ii\up} + W^+[\gamma_{ij\up},\gamma_{ij\dw}] \,,
\end{equation}
where we have used that $W^-[\bm{\gamma}] = U D_{\max}[\bm{\gamma}]$ and
$W^+[\bm{\gamma}] = -UD_{\min}[\bm{\gamma}]$ for $U<0$. An analogous
relation is obtained by exchanging the spin directions.

In periodic systems we may focus on density matrices that comply with
translational symmetry. Using that the eigenvectors of $\bm{\gamma}$
are Bloch states, and in particular for single-band models, the
interaction-energy functional can be regarded as a functional of the
occupation-number distribution $\eta_{\ksig}$. In this case,
\cref{eq:16} implies
\begin{equation}
  \label{eq:17}
  W^-[\eta_{\bm{k}\up}, 1-\eta_{\bm{k}\dw}] =
  U \sum_{\bm{k}} \eta_{\bm{k}\up} + W^+[\eta_{\bm{k}\up},
  \eta_{\bm{k}\dw}] \,.
\end{equation}
Notice that the approximations to $W$ proposed in \cref{eq:11} for the
attractive Hubbard model and in Ref.~\cite{Muller_jul2018} for the
repulsive case fulfill this exact relation in their common domain of
validity, i.e., for $N_{\up}=N_{\dw}=N_a/2$. The \cref{eq:16,eq:17}
can also be used to share and contrast any developments in the
treatment of attractive and repulsive correlations in the Hubbard
model within \gls{ldft}. One may furthermore note that the mapping
between the kets $\ket{\Psi}$ and $\ket{\Psi^h}$ can be extended,
using the same electron-hole transformation, to a mapping between the
mixed states $\hat{\rho}$ and $\hat{\rho}^h$ which can be constructed
by incoherently superposing them. This implies that the relations
\labelcref{eq:12,eq:13,eq:15} between $\gamma_{ij\sigma}$ and
$\gamma_{ij\sigma}^h$ or between $D$ and $D^h$ also hold when the
averages are performed using the corresponding $\hat{\rho}$ and
$\hat{\rho}^h$. Therefore, the relations \labelcref{eq:16,eq:17}
between the attractive- and repulsive-interaction functionals also
apply to the more general set of ensemble-representable density
matrices $\gamma_{ij\sigma}$.

The ground-state energy $E_0$ and occupation-number distribution
$\bm{\eta}^0$ is obtained by minimizing the corresponding energy
functional
\begin{equation}
  \label{eq:18}
  E[\bm{\eta}] = \sum_{\ksig} \veps_{\bm{k}} \, \eta_{\ksig} +
  W[\bm{\eta}] 
\end{equation}
under the constraint $N_{\sigma} = \sum_{\bm{k}} \eta_{\ksig}$ on the
number of spin-$\sigma$ fermions. Let us recall that any \gls{ife}
approximation of the form $W[\bm{\eta}] = W(S[\bm{\eta}])$ (i.e., one
in which $W$ depends on $\bm{\eta}$ through $S$) leads to ground-state
occupation numbers $\eta_{\ksig}^0$ which follow a Fermi-Dirac
function with an effective temperature
$T_{\eff} = -\partial W/\partial S$ that depends only on $U$ and
$N_{\sigma}$~\cite{Muller_jul2018}. The applications discussed in the
following section show that the thus obtained Fermi-Dirac distributions
are in most cases quite close to the exact ground-state occupation numbers
of the attractive Hubbard model. Consequently, the linear
\gls{ife}-approximation~\eqref{eq:11} is expected to yield a sound
description of the most important ground-state observables in the complete
range of interactions from weak to strong correlations.

\section{Results}%
\label{sec:results}

\begin{figure}[ptb]
  \centering
  \includegraphics[scale=\figscale]{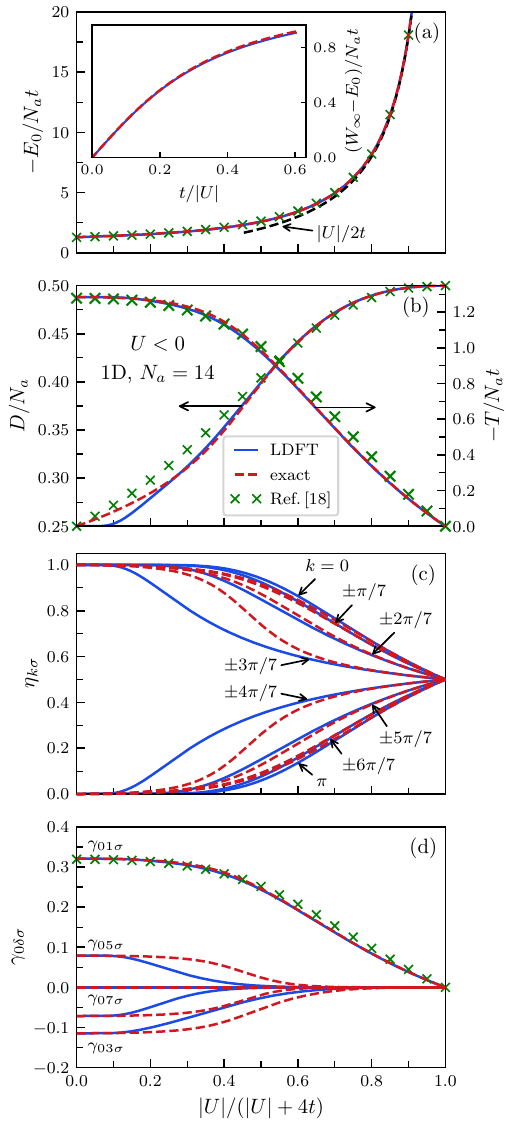}
  \caption{Ground-state properties of the one-dimensional (1D)
    attractive Hubbard model on a 14-site ring with \gls{nn}-hoppings
    $t_{ij} = -t < 0$ and half-band filling as a function of the interaction
    strength~$|U|/t$: (a)~Ground-state energy $E_0$, (b)~average
    number of double occupations $D$ and kinetic energy~$T$,
    (c)~natural-orbital occupation numbers
    $\eta_{\bm{k}\up} = \eta_{\bm{k}\dw}$ and (d)~single-particle
    density-matrix elements
    $\gamma_{0\delta\up} = \gamma_{0\delta\dw}$ between site $i=0$ and
    its $\delta$th \gls{nn}. The linear \gls{ife} approximation
    [\cref{eq:11}, blue curves] is compared with exact numerical
    Lanczos diagonalizations (red dashed curves) and with the results
    of Ref.~\cite{Saubanere_sep2014} whenever available (green
    crosses). The dashed black curve in (a) shows the strongly
    correlated limit of $E_0$ given by $W_{\infty}=-|U|N_a/2$. The
    inset of (a) highlights the energy gain $W_{\oo}-E_0$ in the
    strongly correlated limit ($t/|U|\ll 1$).}%
  \label{fig:2}
\end{figure}

As a first application of the theory we consider the one-dimensional
attractive Hubbard model and compare our results with exact numerical
diagonalizations as well as with previous \gls{ldft}
studies~\cite{Saubanere_sep2014}. In \cref{fig:2} results are given for a
number of relevant ground-state properties of a 14-site ring with
\gls{nn}-hopping $t_{ij} = -t < 0$ and half-band filling, as functions of
the interaction strength $|U|/t$. One observes that the ground-state
energy $E_0$ decreases monotonously with increasing $|U|/t$, since the
pairing energy overcompensates the kinetic-energy increase caused by
the gradual fermion localization. The \gls{ife} results for $E_0$
follow the exact numerical calculations very closely in the complete
range of the interaction strength $|U|/t$ [see \cref{fig:2}\,(a)]. In
the strongly correlated limit, $|U|/t\to\oo$, $E_0$ diverges tending
to $W_{\oo} = -|U|N_a/2$ as $E_0\simeq W_{\oo}-\alpha t^2/|U|$ with
$\alpha>0$, a fact which is highlighted in the inset of
\cref{fig:2}\,(a). The limit $W_{\oo}$ represents the interaction
energy of $N_a/2$~pairs ($N_{\up}=N_{\dw}=N_a/2$) while the additional
energy lowering $-\alpha t^2/|U|$ is a second-order perturbation
correction in the hopping integrals consisting in the virtual breaking
and recombination of pairs. The linear \gls{ife} approximation
reproduces this behavior almost exactly with a leading coefficient
$\alpha_{\IFE} = 2.77$ which is only $0.7\%$ smaller than the exact
value $\alpha_{\ex} = 2.79$ obtained from the Lanczos
diagonalizations. Furthermore, it is interesting to observe that the
accuracy of the \gls{ife} ansatz concerning $E_0$ is not the result of
an important compensation of errors, since the average number of
double occupations~$D$ and the kinetic energy~$T$ are very well
reproduced separately, as shown in \cref{fig:2}\,(b). Only in the
weakly correlated limit ($|U|/t<1$) we find that the present
approximation underestimates the double occupations, which remain too
close to the Hartree-Fock value $D_{\HF}=N_a/4$. This can be ascribed
mainly to deviations from linearity in $W(S)$ for nearly integer
occupation numbers $\eta_{\ksig}\simeq 0$ or $1$, i.e., for small $S$,
which can be seen in \cref{fig:1}. Notice, however, that this does not
affect the accuracy of the ground-state energy $E_0$, since for weak
interactions the kinetic energy, whose functional dependence is
exactly known, largely dominates. In this interaction regime
($|U|/t \lesssim 0.5$) the scaling ansatz proposed in
Ref.~\cite{Saubanere_sep2014} yields a small overestimation of $D$ and
is in general closer to the exact results than the present
calculations. However, as soon as the interaction and hopping
integrals have comparable strengths ($|U|/t \gtrsim 1$) the linear
\gls{ife} approximation clearly outperforms the previous real-space
approach. This concerns not only the double occupations and the
kinetic energy but also the \gls{nn} bond order $\gamma_{01\sigma}$
shown in \cref{fig:2}\,(d).

In \cref{fig:2}\,(c) the Bloch-state occupation numbers
$\eta_{k\sigma}$ are shown as functions of $|U|/t$. One observes that
the present \gls{ife} approximation reproduces the crossover from weak
to strong interactions qualitatively well. For $|U|/t\ll 1$ one
obtains $\eta_{k\sigma}=1$ for $|k|<\pi/2$ and $\eta_{k\sigma}=0$
otherwise, as expected for weakly interacting fermions. As the
interaction strength increases, charge fluctuations are progressively
suppressed in order to favor local pairing. Consequently,
$\eta_{k\sigma}$ decreases (increases) for $|k|<\pi/2$ ($k>\pi/2$)
until $\eta_{k\sigma}\to 1/2$ for all $k\sigma$ in the strongly
correlated limit. Quantitatively, the comparison with the exact
results is in most cases remarkably good. Only for $|k|=3\pi/7$
($|k|=4\pi/7$) at intermediate interaction strength we find that
the \gls{ife} ansatz underestimates (overestimates) $\eta_{k\sigma}$
significantly. This means that in these cases the excitations of electrons
across the Fermi energy $\veps_F=0$, which are the result of
correlations, are overestimated. The same behavior has already been
observed in the repulsive case~\cite{Muller_jul2018}, which is not
surprising since the attractive and repulsive Hubbard models are
related by an electron-hole transformation~\footnote{%
  The electron-hole transformation relating positive-$U$ and
  negative-$U$ Hubbard models is obtained by introducing the following
  hole-annihilation operators: $\ca[h]{i\up}=\ca{i\up}$ and
  $\ca[h]{i\dw}=\ct{i\dw}$. It implies a change of sign in the
  down-spin hopping integrals and therefore leaves $\hat{H}$ invariant
  only for bipartite lattices, in which case one sets
  $\ca[h]{i\dw}=-\ct{i\dw}$ for $i$ belonging to one of the two
  sublattices. In addition an irrelevant shift of the total energy is
  obtained. Notice that if $\Delta S_z = N_{\dw} - N_a/2\neq 0$ a
  change in the total spin-polarization $S_z^{(h)} = S_z+\Delta S_z$
  is involved.}. %
In fact, it is easy to show that in the case of bipartite lattices and
half-band filling the ground-state occupation numbers $\eta_{\ksig}$
and the density matrix $\gamma_{ij\sigma}$ of the Hubbard model are
independent of the sign of $U$. Knowing that the \gls{ife}
approximation respects this symmetry, the same trends are
observed. From a local perspective we observe that our approximation
underestimates the delocalization of the electrons beyond first
\glspl{nn}. Indeed, as shown in \cref{fig:2}\,(d),
$\gamma_{0\delta\sigma}$ tends to zero much faster than the exact
result for $\delta=3,5$ and 7. They are significantly underestimated
in absolute value for $1 \lesssim |U|/t \lesssim 6$. For even
$\delta$, we obtain $\gamma_{0\delta\sigma}=0$ for all $|U|/t$ as
required by the electron-hole symmetry. The $|U|/t$ dependence of
$\gamma_{0\delta\sigma}$ shown in \cref{fig:2}\,(d) coincides with the
one observed in the repulsive model~\cite{Muller_jul2018}.

\begin{figure}[ptb]
  \centering
  \includegraphics[scale=\figscale]{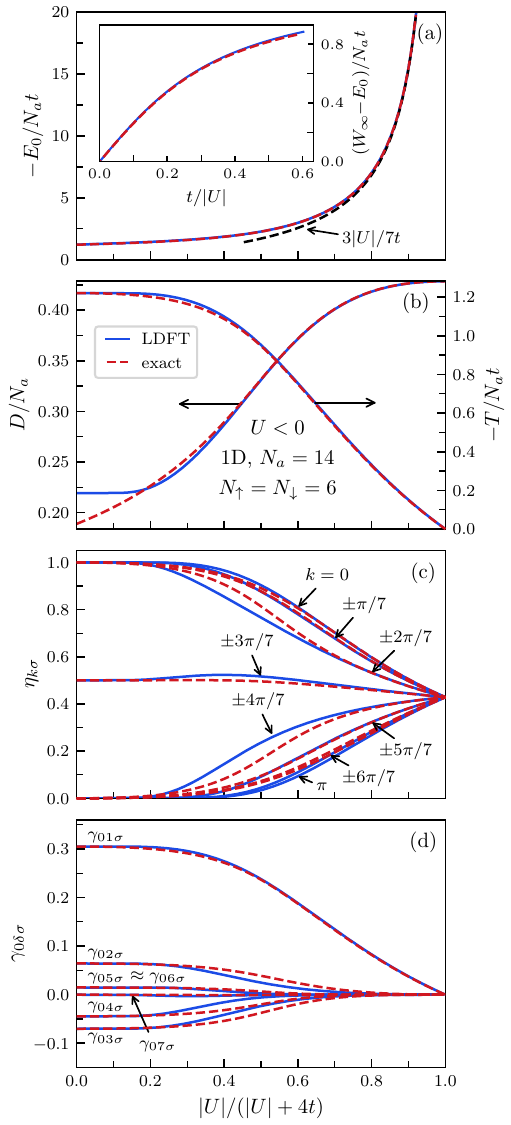}
  \caption{Ground-state properties of the one-dimensional (1D)
    attractive Hubbard model on a 14-site ring with \gls{nn}-hoppings
    $t_{ij} = -t < 0$ and a band filling $N_\sigma/N_a = 3/7$ 
    as a function of the interaction strength~$|U|/t$:
    (a)~Ground-state energy $E_0$, (b)~average
    number of double occupations $D$ and kinetic energy~$T$,
    (c)~natural-orbital occupation numbers
    $\eta_{\bm{k}\up} = \eta_{\bm{k}\dw}$ and (d)~single-particle
    density-matrix elements
    $\gamma_{0\delta\up} = \gamma_{0\delta\dw}$ between site $i=0$ and
    its $\delta$th \gls{nn}. The linear \gls{ife} approximation
    [\cref{eq:11}, blue curves] is compared with exact numerical
    Lanczos diagonalizations (red dashed curves). The dashed black
    curve in (a) shows the strongly
    correlated limit of $E_0$ given by $W_{\infty}=-|U|N_\sigma$. The
    inset of (a) highlights the energy gain $W_{\infty}-E_0$ in the
    strongly correlated limit ($t/|U|\ll 1$).}%
  \label{fig:1D-nhf}
\end{figure}

It is also interesting to assess the accuracy of the IFE ansatz away from
half-band filling. In Fig.~\ref{fig:1D-nhf} exact and IFE results for
several ground-state properties are shown for a 1D ring near 1/4-band
filling, namely, $N_\sigma/N_a = 3/7$. The dependence of the ground-state
properties on the interaction strength is found to be very similar to the
half-filled case. These trends are remarkably well reproduced by the present
approach in the complete range of interaction strength, from weak to strong
correlations. Quantitatively, the IFE results are in some cases slightly more
accurate than for $N_\sigma/N_a = 1$, for example concerning the occupation
numbers $\eta_{k\sigma}$ (compare Figs.~\ref{fig:2} and
\ref{fig:1D-nhf}). The most significant deviations are observed in the
average number of double occupations $D$ in the weakly correlated limit
[Fig.~\ref{fig:1D-nhf}(b)] where $D$ is overestimated by about 15\%
for $|U|\to 0$. As discussed below, this discrepancy can be traced back
to the presence of degeneracies at the Fermi level of the single-particle
spectrum, which allows a reduction of $D$ in the exact correlated state
below the Hartree-Fock value, even for $U \to 0$. These inaccuracies are of
little consequence for the other ground-state properties since the
kinetic energy largely dominates in this limit. 

\begin{figure}[ptb]
  \centering
  \includegraphics[scale=\figscale]{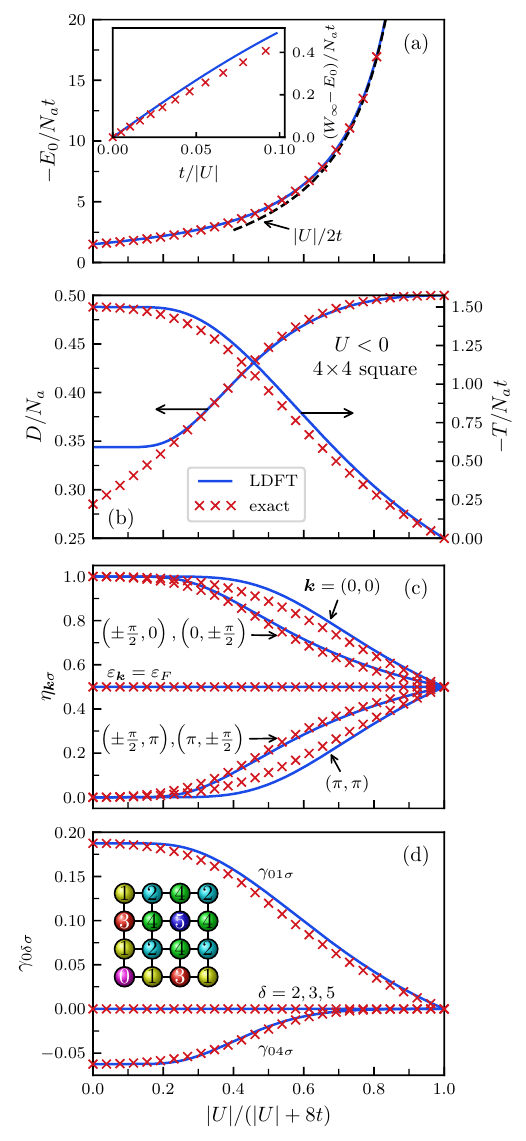}
  \caption{Ground-state properties of the two-dimensional (2D) attractive
    Hubbard model on a 4$\times$4~square-lattice cluster with
    \gls{nn}-hopping $t_{ij} = - t < 0$, periodic boundary conditions and
    half-band filling ($N_{\up} = N_{\dw} = 8$). The linear \gls{ife}
    approximation [\cref{eq:11}] (blue curves) is compared with
    exact numerical Lanczos diagonalizations (red crosses) as a
    function of the interaction strength~$|U|/t$: (a)~ground-state
    energy $E_0$, (b)~average number of double occupations $D$ and
    kinetic energy~$T$, (c)~natural-orbital occupation numbers
    $\eta_{\bm{k}\up} = \eta_{\bm{k}\dw}$ and (d)~density-matrix
    elements $\gamma_{0\delta\up} = \gamma_{0\delta\dw}$ between site
    $i=0$ and its $\delta$th \gls{nn}, as labeled in the inset. The
    dashed black curve in (a) shows the strongly correlated limit of
    $E_0$ given by $W_{\infty}=-|U|N_a/2$. The inset of (a) highlights
    the ground-state energy gain $W_{\infty}-E_0$ in the strongly
    correlated limit ($t/|U|\ll 1$).}%
  \label{fig:3}
\end{figure}

The linear \gls{ife} approximation has also been applied to
two-dimensional (2D) lattices.  In \cref{fig:3} results are shown for
ground-state properties of the half-filled attractive Hubbard model on
a 4$\times$4 square-lattice cluster with \gls{nn} hoppings and
periodic boundary conditions ($N_{\up}=N_{\dw}=8$).  The ground-state
energy $E_0$ obtained with \gls{ldft} follows the exact results very
closely in the complete range of the interaction strength $|U|/t$. The
trends are similar to the 1D case. In particular the strongly
correlated limit, where $E_0\simeq W_{\oo}-\alpha t^2/|U|$, is very
well reproduced with a leading coefficient $\alpha_{\IFE} = 5.55$
that is only $13\%$ larger than the exact value $\alpha_{\ex} = 4.81$
[see the inset of \cref{fig:3}\,(a)].

The \gls{ife} results for the kinetic energy $T$ [\cref{fig:3}\,(b)]
are also very good, although the differences with respect to the exact
diagonalizations are more noticeable than in $E_0$ for intermediate
$|U|/t$. Concerning the average number of double occupations~$D$, one
observes that any significant discrepancies between the \gls{ife} and
exact results are restricted to the weakly correlated limit, namely
for $|U|/t\lesssim 2$ [see \cref{fig:3}\,(b)]. These are most probably
the consequence of the strong degeneracy of the single-particle
spectrum of the 4$\times$4~cluster at the partially filled Fermi level
$\veps_F = 0$. In fact, the Slater determinants constructed with
different degenerate Bloch states having
$\veps_{\bm{k}} = \veps_F = 0$ can be linearly combined in order to
enhance $D$ beyond the Hartree-Fock value $D_{\HF}=N_a/4$, without
causing any kinetic-energy increase.  Thus, the ground-state energy
can be lowered in the limit of small $|U|/t$.  Both the \gls{ldft} and
exact calculations take advantage of this possibility and yield that
all 4 Bloch states at the Fermi level are equally occupied with
$\eta_{\ksig}=1/2$ [compare $D_{\IFE}$ and $D_{\ex}$ with $D_{\HF}=N_a/4$
for $|U|/t\to 0$ in \cref{fig:3}\,(b)]. However, the \gls{ife}
approximation overestimates the ability of the system to enhance $D$
by predicting $D_{\IFE}/N_a\simeq 0.34$ whereas
$D_{\ex}/N_a\simeq 0.29$ for $|U|/t\to 0$. This overestimated value of
$D$ remains nearly constant at finite $|U|/t$ until the \gls{ife}
result matches the exact one for $|U|/t\simeq 2$. Further increase of
the interaction strength yields a monotonous increase of $D$ in very
good quantitative agreement with the exact results. This increase
of~$D$ with increasing $|U|/t$ is accompanied with a decrease of the
absolute value of the kinetic energy~$|T|$ as the fermions begin to
form pairs and localize ($T\le 0$). Finally, as the fully localized
state is approached for $|U|/t\to\oo$, only virtual pair-breaking
remains possible and $T$ vanishes proportionally to $-t^2/|U|$. As
shown in \cref{fig:3}\,(b) all these trends are very well reproduced
by the \gls{ife} approximation to \gls{ldft}.

The dependence of the Bloch-state occupation numbers $\eta_{\ksig}$ on
$|U|/t$ is shown in \cref{fig:3}\,(c). In the non-interacting limit
$\eta_{\ksig} = 1$ ($\eta_{\ksig} = 0$) for $\veps_{\bm{k}} < \veps_F$
($\veps_{\bm{k}} > \veps_F$). The degenerate states at the Fermi level
(i.e., for $\veps_{\bm{k}} = \veps_F = 0$) have all
$\eta_{\ksig} = 1/2$ independent of $|U|/t$. As $|U|/t$ increases, one
observes that $\eta_{\ksig}$ decreases (increases) if
$\veps_{\bm{k}} < \veps_F$ ($\veps_{\bm{k}} > \veps_F$), until
$\eta_{\ksig} = 1/2$ is reached for all $\ksig$ in the
strongly-correlated limit, which corresponds to a fully localized
state.  The occupation numbers~$\eta_{\ksig}$ obtained within the
\gls{ife} approximation are remarkably accurate for all~$\bm{k}\sigma$
in the complete range from weak to strong interactions. Furthermore,
note that the approximation respects all the point-group symmetries
of~$\eta_{\ksig}$ in the reciprocal space, since they are inherited
from the corresponding symmetries of the dispersion
relation~$\veps_{\bm{k}}$.

The very good accuracy of the Bloch-state occupation numbers
$\eta_{\ksig}$ anticipates a comparably good accuracy of the
\gls{spdm}~elements $\gamma_{ij\sigma}$, which are shown in
\cref{fig:3}\,(d). One observes how the non-vanishing
$\gamma_{01\sigma}$ and~$\gamma_{04\sigma}$ between first and fourth
neighbors are gradually suppressed as the interaction strength $|U|/t$
increases. This reflects the correlation-induced suppression of charge
fluctuations and the transition to a localized state as the fermions
condense into localized pairs. One may furthermore notice that the
charge fluctuations at longer distances~$|\gamma_{04\sigma}|$ are
suppressed faster than the fluctuations~$\gamma_{01\sigma}$ between
\glspl{nn}, a trend which is reproduced by the proposed \gls{ife}
approximation.

\begin{figure}[ptb]
  \centering
  \includegraphics[scale=\figscale]{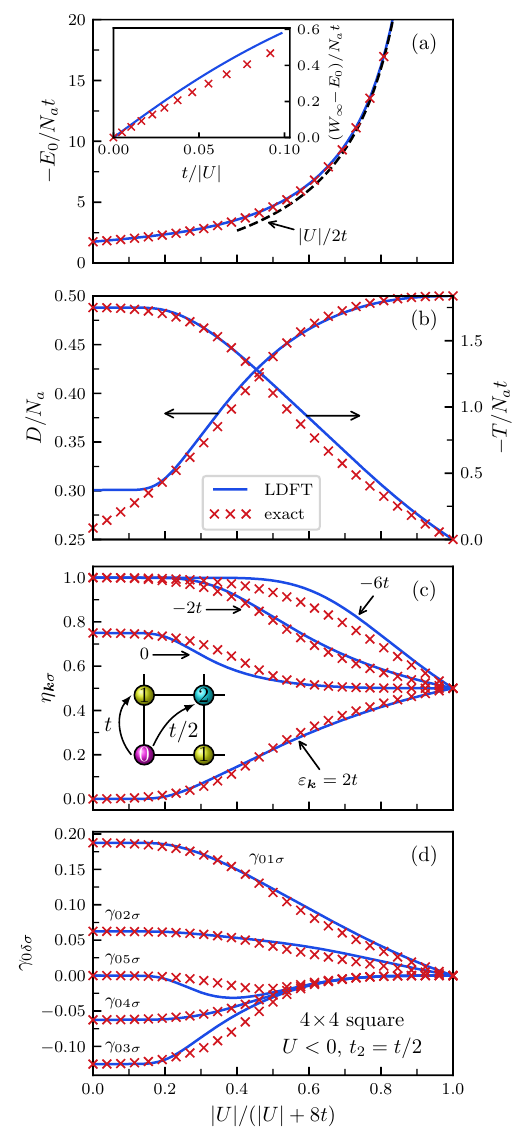}
  \caption{Ground-state properties of the half-filled attractive 2D
    Hubbard model on a 4$\times$4 periodic square lattice with
    \gls{nn} hopping $t_{ij} = -t < 0$ and second-\gls{nn} hopping
    $t_{ij} = -t'= -t/2$.  The
    linear \gls{ife} approximation (blue curves) is compared with
    exact numerical Lanczos diagonalizations (red crosses) as a
    function of the interaction strength $|U|/t$: (a)~ground-state
    energy $E_0$, (b)~average number of double occupations $D$ and
    kinetic energy~$T$, (c)~natural-orbital occupation numbers
    $\eta_{\bm{k}\up} = \eta_{\bm{k}\dw}$ and (d)~density-matrix
    elements $\gamma_{0\delta\up} = \gamma_{0\delta\dw}$ between site
    $i=0$ and its $\delta$th \gls{nn}, as labeled in the inset of
    \cref{fig:3}\,(d). The dashed black curve in (a) shows the
    strongly correlated limit of $E_0$ given by
    $W_{\oo}=-|U|N_a/2$. The inset of (a) highlights the ground-state
    energy gain $W_{\oo}-E_0$ in the strongly correlated limit
    ($t/|U|\ll 1$).}%
  \label{fig:4}
\end{figure}

In \cref{fig:4} the effects of second-\gls{nn} hoppings on the ground-state
properties of the square lattice are investigated by considering a 4$\times$4
cluster with \gls{nn}-hopping $t_{ij} = -t < 0$ and second-\gls{nn}
hoppings $t_{ij} = -t' = -t/2$. Comparison with \cref{fig:3} shows
that the dependencies of the kinetic, interaction and total energies
as functions of $|U|/t$ are not strongly affected by breaking the
bipartite character of the \gls{nn} square lattice.  More significant
changes are observed, as expected, in the density-matrix elements
$\gamma_{ij\sigma}$ and in the occupation numbers $\eta_{\ksig}$ of
the Bloch states [see Figs.~\ref{fig:4}\,(c) and (d)]. The accuracy of
the linear \gls{ife} ansatz remains in general very good when $t'=t/2$
is introduced, for example, concerning the ground-state energy
$E_0$. This is particularly remarkable in the crossover region from
weak to strong correlations ($1\lesssim |U|/t \lesssim 10$) where the
interplay between the fermion delocalization, driven by the
hybridizations, and the tendency to pair formation and localization is
far from trivial. As in the bipartite case ($t'=0$, \cref{fig:3}) the
asymptotic behavior $E_0\simeq -|U|N_a/2-\alpha t^2/|U|$ in the
strongly-correlated limit is qualitatively reproduced as highlighted
in the inset of \cref{fig:4}\,(a). Quantitatively, the coefficient
$\alpha_{\IFE} = 6.93$ is larger than in the case with only first
\gls{nn} hoppings ($\alpha_{\IFE}=5.55$ for $t'=0$) which implies an
increased stabilization due to virtual pair breaking.  This
enhancement of $\alpha$ is in qualitative agreement with the trends
observed in the exact diagonalizations, which yield
$\alpha_{\ex} = 5.65$ for $t'=t/2$ and $\alpha_{\ex}=4.81$ for
$t'=0$. However, in all cases the \gls{ife} functional results in an
overestimation of about $20\%$ of the binding energy.

The average number of ground-state double occupations $D$ and the
kinetic energy $T$ show a dependence on $|U|/t$ that is very similar
to the case where only \gls{nn} hoppings are taken into account
[compare (b) and (c) of Figs.~\ref{fig:3} and~\ref{fig:4}]. The exact
diagonalizations show that for $|U|/t\to 0$ the second-\gls{nn}
hoppings $t'=t/2$ cause a $17\%$ lowering of $T$ at the expense of an
$8\%$ decrease of $D$. Since the kinetic-energy functional is exact in
\gls{ldft}, the above mentioned kinetic-energy lowering is exactly
reproduced by the \gls{ife} approximation. However, $D$ is clearly
overestimated in the weakly-interacting regime. This is probably
related to the degeneracies at the Fermi level of the single-particle
spectrum, as in the $t'=0$ case. Despite the overestimation, the
\gls{ife} calculations yield a $13\%$ reduction of $D$ for
$|U|/t\to 0$ due to the second-\gls{nn} hoppings $t'=t/2$, which is in
good qualitative agreement with the $8\%$ reduction obtained in the
corresponding exact solution. The difficulties found at weak
interactions disappear for $|U|/t \gtrsim 2$ where \gls{ldft} recovers
its usual very good accuracy.

In \cref{fig:4}\,(c) results are shown for the Bloch-state occupation
numbers $\eta_{\ksig}$. As expected, the $\eta_{\ksig}$ which are
larger (smaller) than $1/2$ decrease (increase) as $|U|/t$ increases
reaching the common limit $\eta_{\ksig}=1/2$ for all $\bm{k}$ in the
localized state when $|U|/t\to\oo$. An interesting difference with
respect to the $t'=0$ case is observed at the Fermi energy where
$t'=t/2$ reduces the degeneracy from 6 to 4. Since $\veps_F$ is
occupied by 3 fermions, we have $\eta_{\ksig}=3/4$ in the weakly
correlated limit. Consequently, these occupation numbers decrease with
increasing $|U|/t$, in contrast to the bipartite lattice [see
Figs.~\ref{fig:3}\,(c) and~\ref{fig:4}\,(c)]. Comparison with the
exact diagonalizations shows that the \gls{ife} approximation
reproduces all the Bloch-state occupation numbers with very good
accuracy from weak to strong correlations. Notice, however, that
$\eta_{\ksig}$ is somewhat overestimated for $\bm{k}=(0,0)$
($\veps_{\bm{k}} = -6t$) and intermediate $|U|/t$, which incidentally
explains the overestimation of $|T|$ in this range. For example,
for~$|U|/t=15$, we find that $\eta_{\ksig}$ is overestimated
by~$8.7\%$ for $\bm{k}=(0,0)$ and $|T|$ is overestimated by
about~$17\%$ [see  Figs.~\ref{fig:4}\,(b) and (c)].

The density-matrix elements~$\gamma_{0\delta\sigma}$ between the
lattice site~$i=0$ and its $\delta$th~\gls{nn} are given in
\cref{fig:4}\,(d) as a function of $|U|/t$. The hopping integrals $t'$
between second~\glspl{nn} introduce hybridizations between sites
belonging to the same sublattice (e.g., between $i=0$ and $i=2,3$ and
$5$) which are absent in the bipartite case [see the inset of
\cref{fig:3}\,(d)]. As a result $\gamma_{0\delta\sigma}$ is no longer
zero for $\delta=2,3$ and $5$. In particular $\gamma_{05\sigma}$ shows
a remarkable non-monotonous behavior with a minimum at
$|U|/t\simeq 8$. This is qualitatively reproduced by the \gls{ife}
ansatz, although somewhat exaggerated and slightly shifted to smaller
$|U|/t$. This behavior might be related to the rapid decrease of
$\eta_{\ksig}$ for $\veps_{\bm{k}}=\veps_F=0$ in the same range of
$|U|/t$ [see \cref{fig:4}\,(c)].

\begin{figure}[ptb]
  \centering
  \includegraphics[scale=\figscale]{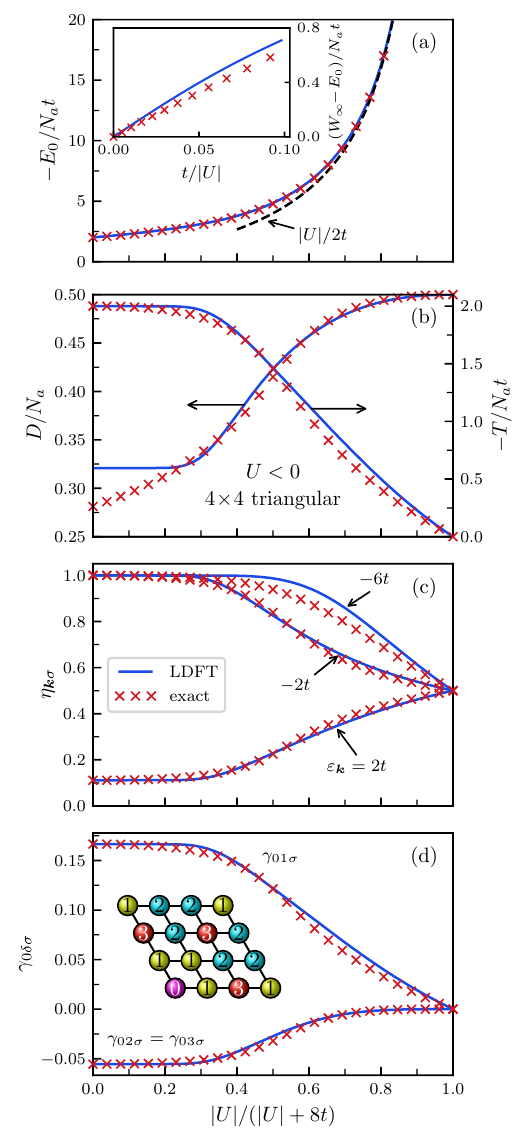}
  \caption{Ground-state properties of the two-dimensional (2D) attractive
    Hubbard model on a 4$\times$4 triangular-lattice cluster with \gls{nn}
    hopping $t_{ij} = -t < 0$, periodic boundary conditions and half-band
    filling ($N_{\up}=N_{\dw}=8$). The linear \gls{ife}
    approximation (blue curves) is compared with exact numerical
    Lanczos diagonalizations (red crosses) as a function of the
    interaction strength $|U|/t$: (a)~ground-state energy $E_0$,
    (b)~average number of double occupations $D$ and kinetic
    energy~$T$, (c)~natural-orbital occupation numbers
    $\eta_{\bm{k}\up} = \eta_{\bm{k}\dw}$ and (d)~density-matrix
    elements $\gamma_{0\delta\up} = \gamma_{0\delta\dw}$ between site
    $i=0$ and its $\delta$th \gls{nn}, as labeled in the inset. The
    dashed black curve in (a) shows the strongly correlated limit of
    $E_0$ given by $W_{\oo}=-|U|N_a/2$. The inset of (a) highlights
    the ground-state energy gain $W_{\oo}-E_0$ in the strongly
    correlated limit ($t/U\ll 1$).}%
  \label{fig:5}
\end{figure}

As an example of a non-bipartite lattice with triangular \gls{nn}
loops we consider the 2D triangular lattice, which is modeled by a
4$\times$4 cluster with periodic boundary conditions [see the inset of
\cref{fig:5}\,(d)]. As in previous cases, the comparison with exact
results shows that the \gls{ife} approximation gives a very accurate
account of the ground-state energy~$E_0$ for all $|U|/t$ [see
\cref{fig:5}\,(a)]. In the limit of strong correlations in particular
we obtain the right asymptotic dependence
$E_0 \simeq -|U|N_a/2 - \alpha t^2/|U|$ with
$\alpha_{\mathrm{IFE}} = 8.32$. This should be compared with the exact
leading correction having $\alpha_{\mathrm{ex}} = 6.73$ as shown in
the inset of \cref{fig:5}\,(a). Also the kinetic energy~$T$ is very
well reproduced, which implies that the accuracy of $E_0$ is not the
result of strong compensation of errors in the kinetic and interaction
contributions [see \cref{fig:5}\,(b)]. Only the average number of
double occupations in the weakly interacting regime
($|U|/t\lesssim 3$) shows, as in previous cases, a significant
overestimation which can be traced back to the degeneracies in the
single-particle spectrum at the Fermi level.

The Bloch-state occupation numbers~$\eta_{\ksig}$ given in
\cref{fig:5}\,(c) display the familiar trend as a function of $|U|/t$
which ends up in localization in the strongly correlated limit (i.e.,
$\eta_{\ksig}=1/2$ for all~$\ksig$ at $|U|/t\to\oo$). At half-band
filling the Fermi energy $\veps_F$ is given by the highest
single-particle level $\veps_{\bm{k}}=2t$, which is 9-fold degenerate.
For $|U|/t=0$ one thus finds $\eta_{\ksig} = 1/9$ for all the
degenerate Bloch states since $\veps_F$ is occupied by only one
fermion per spin. The comparison with the exact numerical results
shows that the \gls{ife} approximation yields a very accurate
occupation-number distribution $\eta_{\ksig}$ for all $|U|/t$. Only
the occupation of the lowest-lying Bloch state having $\bm{k}=(0,0)$
and $\veps_{\bm{k}} = -6t$ is somewhat overestimated for intermediate
values $|U|/t$.  Finally, the very good results for $\eta_{\ksig}$
also explain the high accuracy of the \gls{ife} approximation for the
density-matrix elements~$\gamma_{0\delta\sigma}$ shown in
\cref{fig:5}\,(d).

\section{Conclusion}%
\label{sec:summary}
\glsresetall{}

An interaction-energy functional has been developed in order to
investigate the ground state of the Hubbard model with attractive
pairing interactions in the framework of \gls{ldft}. Our approach
takes a reciprocal $\bm{k}$-space perspective and exploits a newly
revealed approximate functional relation between the interaction
energy $W[\bm{\eta}]$ of the Hubbard model corresponding to the
Bloch-state occupation-number distribution $\eta_{\ksig}$ and the
entropy $S[\bm{\eta}]$ of a system of non-interacting fermions having
the same occupation numbers $\eta_{\ksig}$. This has opened up a new
perspective to the ground-state problem of periodic systems, which
takes into account the dependence of the central interaction-energy
functional $W[\bm{\gamma}]$ on all elements of the
\gls{spdm}~$\bm{\gamma}$ and thus leverages the universality of
\gls{ldft}.  The relation between $W[\bm{\eta}]$ and $S[\bm{\eta}]$
has been shown to be approximately linear for a wide range of
ground-state representable occupation-number distributions
$\eta_{\ksig}$. On this basis a simple and very effective
approximation to $W[\bm{\eta}]$ has been inferred.  The flexibility
and efficacy of this linear \acrlong{ife} ansatz has been demonstrated
in applications to the attractive 1D and 2D Hubbard model on bipartite
and non-bipartite lattices.  Comparisons with exact numerical Lanczos
diagonalizations on finite clusters have demonstrated the remarkable
accuracy of this approximation in the complete attractive-interaction
range from weak to strong correlations. The present formulation, which
incorporates a statistical or information-theory perspective, turns
out to open a useful alternative route in the search for increasingly
accurate and more broadly applicable approximations to the interaction
energy of strongly correlated systems in the framework of \gls{ldft}.

\section{Acknowledgements}
Helpful discussions with W.\ Töws are gratefully
acknowledged. Computer resources were provided by the IT Service
Center of the University of Kassel and by the Center for Scientific
Computing of the Goethe University of Frankfurt.

\bibliographystyle{files/bibstyle}
\bibliography{files/bibliography}

\begin{thebibliography}{47}%
\makeatletter
\providecommand \@ifxundefined [1]{%
 \@ifx{#1\undefined}
}%
\providecommand \@ifnum [1]{%
 \ifnum #1\expandafter \@firstoftwo
 \else \expandafter \@secondoftwo
 \fi
}%
\providecommand \@ifx [1]{%
 \ifx #1\expandafter \@firstoftwo
 \else \expandafter \@secondoftwo
 \fi
}%
\providecommand \natexlab [1]{#1}%
\providecommand \enquote  [1]{``#1''}%
\providecommand \bibnamefont  [1]{#1}%
\providecommand \bibfnamefont [1]{#1}%
\providecommand \citenamefont [1]{#1}%
\providecommand \href@noop [0]{\@secondoftwo}%
\providecommand \href [0]{\begingroup \@sanitize@url \@href}%
\providecommand \@href[1]{\@@startlink{#1}\@@href}%
\providecommand \@@href[1]{\endgroup#1\@@endlink}%
\providecommand \@sanitize@url [0]{\catcode `\\12\catcode `\$12\catcode
  `\&12\catcode `\#12\catcode `\^12\catcode `\_12\catcode `\%12\relax}%
\providecommand \@@startlink[1]{}%
\providecommand \@@endlink[0]{}%
\providecommand \url  [0]{\begingroup\@sanitize@url \@url }%
\providecommand \@url [1]{\endgroup\@href {#1}{\urlprefix }}%
\providecommand \urlprefix  [0]{URL }%
\providecommand \Eprint [0]{\href }%
\providecommand \doibase [0]{http://dx.doi.org/}%
\providecommand \selectlanguage [0]{\@gobble}%
\providecommand \bibinfo  [0]{\@secondoftwo}%
\providecommand \bibfield  [0]{\@secondoftwo}%
\providecommand \translation [1]{[#1]}%
\providecommand \BibitemOpen [0]{}%
\providecommand \bibitemStop [0]{}%
\providecommand \bibitemNoStop [0]{.\EOS\space}%
\providecommand \EOS [0]{\spacefactor3000\relax}%
\providecommand \BibitemShut  [1]{\csname bibitem#1\endcsname}%
\let\auto@bib@innerbib\@empty
\bibitem [{\citenamefont {Dagotto}(1994)}]{Dagotto_jul1994}%
  \BibitemOpen
  \bibfield  {author} {\bibinfo {author} {\bibfnamefont {E.}~\bibnamefont
  {Dagotto}},\ }\bibfield  {title} {\enquote {\bibinfo {title} {Correlated
  electrons in high-temperature superconductors},}\ }\href {\doibase
  10.1103/RevModPhys.66.763} {\bibfield  {journal} {\bibinfo  {journal} {Rev.
  Mod. Phys.}\ }\textbf {\bibinfo {volume} {66}},\ \bibinfo {pages} {763}
  (\bibinfo {year} {1994})}\BibitemShut {NoStop}%
\bibitem [{\citenamefont {Bednorz}\ and\ \citenamefont
  {Müller}(1986)}]{Bednorz_jun1986}%
  \BibitemOpen
  \bibfield  {author} {\bibinfo {author} {\bibfnamefont {J.~G.}\ \bibnamefont
  {Bednorz}}\ and\ \bibinfo {author} {\bibfnamefont {K.~A.}\ \bibnamefont
  {Müller}},\ }\bibfield  {title} {\enquote {\bibinfo {title} {Possible
  high{$T_c$} superconductivity in the {Ba-La-Cu-O} system},}\ }\href {\doibase
  10.1007/bf01303701} {\bibfield  {journal} {\bibinfo  {journal} {Z. Physik B -
  Condensed Matter}\ }\textbf {\bibinfo {volume} {64}},\ \bibinfo {pages} {189}
  (\bibinfo {year} {1986})}\BibitemShut {NoStop}%
\bibitem [{\citenamefont {Bardeen}\ \emph {et~al.}(1957)\citenamefont
  {Bardeen}, \citenamefont {Cooper},\ and\ \citenamefont
  {Schrieffer}}]{Bardeen_1957}%
  \BibitemOpen
  \bibfield  {author} {\bibinfo {author} {\bibfnamefont {J.}~\bibnamefont
  {Bardeen}}, \bibinfo {author} {\bibfnamefont {L.~N.}\ \bibnamefont {Cooper}},
  \ and\ \bibinfo {author} {\bibfnamefont {J.~R.}\ \bibnamefont {Schrieffer}},\
  }\bibfield  {title} {\enquote {\bibinfo {title} {Theory of
  {{Superconductivity}}},}\ }\href {\doibase 10.1103/PhysRev.108.1175}
  {\bibfield  {journal} {\bibinfo  {journal} {Phys. Rev.}\ }\textbf {\bibinfo
  {volume} {108}},\ \bibinfo {pages} {1175} (\bibinfo {year}
  {1957})}\BibitemShut {NoStop}%
\bibitem [{\citenamefont {Bickers}\ \emph {et~al.}(1987)\citenamefont
  {Bickers}, \citenamefont {Scalapino},\ and\ \citenamefont
  {Scalettar}}]{Bickers_aug1987}%
  \BibitemOpen
  \bibfield  {author} {\bibinfo {author} {\bibfnamefont {N.}~\bibnamefont
  {Bickers}}, \bibinfo {author} {\bibfnamefont {D.}~\bibnamefont {Scalapino}},
  \ and\ \bibinfo {author} {\bibfnamefont {R.}~\bibnamefont {Scalettar}},\
  }\bibfield  {title} {\enquote {\bibinfo {title} {{CDW} and {SDW} mediated
  pairing interactions},}\ }\href {\doibase 10.1142/s0217979287001079}
  {\bibfield  {journal} {\bibinfo  {journal} {Int. J. Mod. Phys. B}\ }\textbf
  {\bibinfo {volume} {01}},\ \bibinfo {pages} {687} (\bibinfo {year}
  {1987})}\BibitemShut {NoStop}%
\bibitem [{\citenamefont {Baskaran}\ and\ \citenamefont
  {Anderson}(1988)}]{Baskaran_jan1988}%
  \BibitemOpen
  \bibfield  {author} {\bibinfo {author} {\bibfnamefont {G.}~\bibnamefont
  {Baskaran}}\ and\ \bibinfo {author} {\bibfnamefont {P.~W.}\ \bibnamefont
  {Anderson}},\ }\bibfield  {title} {\enquote {\bibinfo {title} {Gauge theory
  of high-temperature superconductors and strongly correlated fermi systems},}\
  }\href {\doibase 10.1103/PhysRevB.37.580} {\bibfield  {journal} {\bibinfo
  {journal} {Phys. Rev. B}\ }\textbf {\bibinfo {volume} {37}},\ \bibinfo
  {pages} {580} (\bibinfo {year} {1988})}\BibitemShut {NoStop}%
\bibitem [{\citenamefont {Kotliar}\ and\ \citenamefont
  {Liu}(1988)}]{Kotliar_sep1988}%
  \BibitemOpen
  \bibfield  {author} {\bibinfo {author} {\bibfnamefont {G.}~\bibnamefont
  {Kotliar}}\ and\ \bibinfo {author} {\bibfnamefont {J.}~\bibnamefont {Liu}},\
  }\bibfield  {title} {\enquote {\bibinfo {title} {Superexchange mechanism and
  d-wave superconductivity},}\ }\href {\doibase 10.1103/PhysRevB.38.5142}
  {\bibfield  {journal} {\bibinfo  {journal} {Phys. Rev. B}\ }\textbf {\bibinfo
  {volume} {38}},\ \bibinfo {pages} {5142} (\bibinfo {year}
  {1988})}\BibitemShut {NoStop}%
\bibitem [{\citenamefont {Schrieffer}\ \emph {et~al.}(1989)\citenamefont
  {Schrieffer}, \citenamefont {Wen},\ and\ \citenamefont
  {Zhang}}]{Schrieffer_jun1989}%
  \BibitemOpen
  \bibfield  {author} {\bibinfo {author} {\bibfnamefont {J.~R.}\ \bibnamefont
  {Schrieffer}}, \bibinfo {author} {\bibfnamefont {X.~G.}\ \bibnamefont {Wen}},
  \ and\ \bibinfo {author} {\bibfnamefont {S.~C.}\ \bibnamefont {Zhang}},\
  }\bibfield  {title} {\enquote {\bibinfo {title} {Dynamic spin fluctuations
  and the bag mechanism of high-${T}_{c}$ superconductivity},}\ }\href
  {\doibase 10.1103/PhysRevB.39.11663} {\bibfield  {journal} {\bibinfo
  {journal} {Phys. Rev. B}\ }\textbf {\bibinfo {volume} {39}},\ \bibinfo
  {pages} {11663} (\bibinfo {year} {1989})}\BibitemShut {NoStop}%
\bibitem [{\citenamefont {Kampf}\ and\ \citenamefont
  {Schrieffer}(1990)}]{Kampf_apr1990}%
  \BibitemOpen
  \bibfield  {author} {\bibinfo {author} {\bibfnamefont {A.}~\bibnamefont
  {Kampf}}\ and\ \bibinfo {author} {\bibfnamefont {J.~R.}\ \bibnamefont
  {Schrieffer}},\ }\bibfield  {title} {\enquote {\bibinfo {title} {Pseudogaps
  and the spin-bag approach to high-${T}_{c}$ superconductivity},}\ }\href
  {\doibase 10.1103/PhysRevB.41.6399} {\bibfield  {journal} {\bibinfo
  {journal} {Phys. Rev. B}\ }\textbf {\bibinfo {volume} {41}},\ \bibinfo
  {pages} {6399} (\bibinfo {year} {1990})}\BibitemShut {NoStop}%
\bibitem [{\citenamefont {Anderson}(1991)}]{Anderson_1991}%
  \BibitemOpen
  \bibfield  {author} {\bibinfo {author} {\bibfnamefont {P.~W.}\ \bibnamefont
  {Anderson}},\ }\bibfield  {title} {\enquote {\bibinfo {title} {{The Theory of
  High $T_c$ Superconductivity}},}\ }in\ \href {\doibase
  10.1007/978-1-4615-3338-2_1} {\emph {\bibinfo {booktitle} {High-Temperature
  Superconductivity}}}\ (\bibinfo  {publisher} {Springer {US}},\ \bibinfo
  {year} {1991})\ pp.\ \bibinfo {pages} {1--6}\BibitemShut {NoStop}%
\bibitem [{\citenamefont {Anderson}\ and\ \citenamefont
  {Bedell}(1998)}]{Anderson_jul1998}%
  \BibitemOpen
  \bibfield  {author} {\bibinfo {author} {\bibfnamefont {P.~W.}\ \bibnamefont
  {Anderson}}\ and\ \bibinfo {author} {\bibfnamefont {K.~S.}\ \bibnamefont
  {Bedell}},\ }\bibfield  {title} {\enquote {\bibinfo {title} {{The Theory of
  Superconductivity in the High-$T_c$ Cuprates}},}\ }\href {\doibase
  10.1063/1.882300} {\bibfield  {journal} {\bibinfo  {journal} {Phys. Today}\
  }\textbf {\bibinfo {volume} {51}},\ \bibinfo {pages} {64} (\bibinfo {year}
  {1998})}\BibitemShut {NoStop}%
\bibitem [{\citenamefont {Hubbard}(1963)}]{Hubbard_nov1963}%
  \BibitemOpen
  \bibfield  {author} {\bibinfo {author} {\bibfnamefont {J.}~\bibnamefont
  {Hubbard}},\ }\bibfield  {title} {\enquote {\bibinfo {title} {Electron
  {Correlations} in {Narrow Energy Bands}},}\ }\href {\doibase
  10.1098/rspa.1963.0204} {\bibfield  {journal} {\bibinfo  {journal} {Proc. R.
  Soc. Lond. A}\ }\textbf {\bibinfo {volume} {276}},\ \bibinfo {pages} {238}
  (\bibinfo {year} {1963})}\BibitemShut {NoStop}%
\bibitem [{\citenamefont {Kanamori}(1963)}]{Kanamori_jan1963}%
  \BibitemOpen
  \bibfield  {author} {\bibinfo {author} {\bibfnamefont {J.}~\bibnamefont
  {Kanamori}},\ }\bibfield  {title} {\enquote {\bibinfo {title} {Electron
  {Correlation} and {Ferromagnetism} of {Transition Metals}},}\ }\href
  {\doibase 10.1143/PTP.30.275} {\bibfield  {journal} {\bibinfo  {journal}
  {Prog. Theor. Phys.}\ }\textbf {\bibinfo {volume} {30}},\ \bibinfo {pages}
  {275} (\bibinfo {year} {1963})}\BibitemShut {NoStop}%
\bibitem [{\citenamefont {Gutzwiller}(1963)}]{Gutzwiller_mar1963}%
  \BibitemOpen
  \bibfield  {author} {\bibinfo {author} {\bibfnamefont {M.~C.}\ \bibnamefont
  {Gutzwiller}},\ }\bibfield  {title} {\enquote {\bibinfo {title} {Effect of
  {Correlation} on the {Ferromagnetism} of {Transition Metals}},}\ }\href
  {\doibase 10.1103/PhysRevLett.10.159} {\bibfield  {journal} {\bibinfo
  {journal} {Phys. Rev. Lett.}\ }\textbf {\bibinfo {volume} {10}},\ \bibinfo
  {pages} {159} (\bibinfo {year} {1963})}\BibitemShut {NoStop}%
\bibitem [{\citenamefont {Hirsch}\ and\ \citenamefont
  {Scalapino}(1985)}]{Hirsch_jul1985}%
  \BibitemOpen
  \bibfield  {author} {\bibinfo {author} {\bibfnamefont {J.~E.}\ \bibnamefont
  {Hirsch}}\ and\ \bibinfo {author} {\bibfnamefont {D.~J.}\ \bibnamefont
  {Scalapino}},\ }\bibfield  {title} {\enquote {\bibinfo {title} {{Excitonic
  mechanism for superconductivity in a quasi-one-dimensional system}},}\ }\href
  {\doibase 10.1103/PhysRevB.32.117} {\bibfield  {journal} {\bibinfo  {journal}
  {Phys. Rev. B}\ }\textbf {\bibinfo {volume} {32}},\ \bibinfo {pages} {117}
  (\bibinfo {year} {1985})}\BibitemShut {NoStop}%
\bibitem [{\citenamefont {Hirsch}\ and\ \citenamefont
  {Scalapino}(1986)}]{Hirsch_jun1986}%
  \BibitemOpen
  \bibfield  {author} {\bibinfo {author} {\bibfnamefont {J.~E.}\ \bibnamefont
  {Hirsch}}\ and\ \bibinfo {author} {\bibfnamefont {D.~J.}\ \bibnamefont
  {Scalapino}},\ }\bibfield  {title} {\enquote {\bibinfo {title} {{Enhanced
  Superconductivity in Quasi Two-Dimensional Systems}},}\ }\href {\doibase
  10.1103/PhysRevLett.56.2732} {\bibfield  {journal} {\bibinfo  {journal}
  {Phys. Rev. Lett.}\ }\textbf {\bibinfo {volume} {56}},\ \bibinfo {pages}
  {2732} (\bibinfo {year} {1986})}\BibitemShut {NoStop}%
\bibitem [{\citenamefont {Scalettar}\ \emph {et~al.}(1989)\citenamefont
  {Scalettar}, \citenamefont {Loh}, \citenamefont {Gubernatis}, \citenamefont
  {Moreo}, \citenamefont {White}, \citenamefont {Scalapino}, \citenamefont
  {Sugar},\ and\ \citenamefont {Dagotto}}]{Scalettar_mar1989}%
  \BibitemOpen
  \bibfield  {author} {\bibinfo {author} {\bibfnamefont {R.~T.}\ \bibnamefont
  {Scalettar}}, \bibinfo {author} {\bibfnamefont {E.~Y.}\ \bibnamefont {Loh}},
  \bibinfo {author} {\bibfnamefont {J.~E.}\ \bibnamefont {Gubernatis}},
  \bibinfo {author} {\bibfnamefont {A.}~\bibnamefont {Moreo}}, \bibinfo
  {author} {\bibfnamefont {S.~R.}\ \bibnamefont {White}}, \bibinfo {author}
  {\bibfnamefont {D.~J.}\ \bibnamefont {Scalapino}}, \bibinfo {author}
  {\bibfnamefont {R.~L.}\ \bibnamefont {Sugar}}, \ and\ \bibinfo {author}
  {\bibfnamefont {E.}~\bibnamefont {Dagotto}},\ }\bibfield  {title} {\enquote
  {\bibinfo {title} {{Phase diagram of the two-dimensional negative-$U$ Hubbard
  model}},}\ }\href {\doibase 10.1103/PhysRevLett.62.1407} {\bibfield
  {journal} {\bibinfo  {journal} {Phys. Rev. Lett.}\ }\textbf {\bibinfo
  {volume} {62}},\ \bibinfo {pages} {1407} (\bibinfo {year}
  {1989})}\BibitemShut {NoStop}%
\bibitem [{\citenamefont {Paiva}\ \emph {et~al.}(2004)\citenamefont {Paiva},
  \citenamefont {dos Santos}, \citenamefont {Scalettar},\ and\ \citenamefont
  {Denteneer}}]{Paiva_may2004}%
  \BibitemOpen
  \bibfield  {author} {\bibinfo {author} {\bibfnamefont {T.}~\bibnamefont
  {Paiva}}, \bibinfo {author} {\bibfnamefont {R.~R.}\ \bibnamefont {dos
  Santos}}, \bibinfo {author} {\bibfnamefont {R.~T.}\ \bibnamefont
  {Scalettar}}, \ and\ \bibinfo {author} {\bibfnamefont {P.~J.~H.}\
  \bibnamefont {Denteneer}},\ }\bibfield  {title} {\enquote {\bibinfo {title}
  {{Critical temperature for the two-dimensional attractive Hubbard model}},}\
  }\href {\doibase 10.1103/PhysRevB.69.184501} {\bibfield  {journal} {\bibinfo
  {journal} {Phys. Rev. B}\ }\textbf {\bibinfo {volume} {69}},\ \bibinfo
  {pages} {184501} (\bibinfo {year} {2004})}\BibitemShut {NoStop}%
\bibitem [{\citenamefont {Saubanère}\ and\ \citenamefont
  {Pastor}(2014)}]{Saubanere_sep2014}%
  \BibitemOpen
  \bibfield  {author} {\bibinfo {author} {\bibfnamefont {M.}~\bibnamefont
  {Saubanère}}\ and\ \bibinfo {author} {\bibfnamefont {G.~M.}\ \bibnamefont
  {Pastor}},\ }\bibfield  {title} {\enquote {\bibinfo {title} {Lattice
  density-functional theory of the attractive {Hubbard} model},}\ }\href
  {\doibase 10.1103/PhysRevB.90.125128} {\bibfield  {journal} {\bibinfo
  {journal} {Phys. Rev. B}\ }\textbf {\bibinfo {volume} {90}},\ \bibinfo
  {pages} {125128} (\bibinfo {year} {2014})}\BibitemShut {NoStop}%
\bibitem [{\citenamefont {Stoner}(1938)}]{Stoner_apr1938}%
  \BibitemOpen
  \bibfield  {author} {\bibinfo {author} {\bibfnamefont {E.~C.}\ \bibnamefont
  {Stoner}},\ }\bibfield  {title} {\enquote {\bibinfo {title} {Collective
  electron ferronmagnetism},}\ }\href {\doibase 10.1098/rspa.1938.0066}
  {\bibfield  {journal} {\bibinfo  {journal} {Proc. R. Soc. Lond. A}\ }\textbf
  {\bibinfo {volume} {165}},\ \bibinfo {pages} {372} (\bibinfo {year}
  {1938})}\BibitemShut {NoStop}%
\bibitem [{\citenamefont {Anderson}(1961)}]{Anderson_oct1961}%
  \BibitemOpen
  \bibfield  {author} {\bibinfo {author} {\bibfnamefont {P.~W.}\ \bibnamefont
  {Anderson}},\ }\bibfield  {title} {\enquote {\bibinfo {title} {Localized
  {Magnetic States} in {Metals}},}\ }\href {\doibase 10.1103/PhysRev.124.41}
  {\bibfield  {journal} {\bibinfo  {journal} {Phys. Rev.}\ }\textbf {\bibinfo
  {volume} {124}},\ \bibinfo {pages} {41} (\bibinfo {year} {1961})}\BibitemShut
  {NoStop}%
\bibitem [{\citenamefont {Kondo}(1964)}]{Kondo_jul1964}%
  \BibitemOpen
  \bibfield  {author} {\bibinfo {author} {\bibfnamefont {J.}~\bibnamefont
  {Kondo}},\ }\bibfield  {title} {\enquote {\bibinfo {title} {{Resistance
  Minimum in Dilute Magnetic Alloys}},}\ }\href {\doibase 10.1143/ptp.32.37}
  {\bibfield  {journal} {\bibinfo  {journal} {Prog. Theor. Phys.}\ }\textbf
  {\bibinfo {volume} {32}},\ \bibinfo {pages} {37} (\bibinfo {year}
  {1964})}\BibitemShut {NoStop}%
\bibitem [{\citenamefont {Micnas}\ \emph {et~al.}(1990)\citenamefont {Micnas},
  \citenamefont {Ranninger},\ and\ \citenamefont
  {Robaszkiewicz}}]{Micnas_jan1990}%
  \BibitemOpen
  \bibfield  {author} {\bibinfo {author} {\bibfnamefont {R.}~\bibnamefont
  {Micnas}}, \bibinfo {author} {\bibfnamefont {J.}~\bibnamefont {Ranninger}}, \
  and\ \bibinfo {author} {\bibfnamefont {S.}~\bibnamefont {Robaszkiewicz}},\
  }\bibfield  {title} {\enquote {\bibinfo {title} {Superconductivity in
  narrow-band systems with local nonretarded attractive interactions},}\ }\href
  {\doibase 10.1103/RevModPhys.62.113} {\bibfield  {journal} {\bibinfo
  {journal} {Rev. Mod. Phys.}\ }\textbf {\bibinfo {volume} {62}},\ \bibinfo
  {pages} {113} (\bibinfo {year} {1990})}\BibitemShut {NoStop}%
\bibitem [{\citenamefont {Marsiglio}(1997)}]{Marsiglio_jan1997}%
  \BibitemOpen
  \bibfield  {author} {\bibinfo {author} {\bibfnamefont {F.}~\bibnamefont
  {Marsiglio}},\ }\bibfield  {title} {\enquote {\bibinfo {title} {{Evaluation
  of the BCS approximation for the attractive Hubbard model in one
  dimension}},}\ }\href {\doibase 10.1103/PhysRevB.55.575} {\bibfield
  {journal} {\bibinfo  {journal} {Phys. Rev. B}\ }\textbf {\bibinfo {volume}
  {55}},\ \bibinfo {pages} {575} (\bibinfo {year} {1997})}\BibitemShut
  {NoStop}%
\bibitem [{\citenamefont {Tanaka}\ and\ \citenamefont
  {Marsiglio}(1999)}]{Tanaka_aug1999}%
  \BibitemOpen
  \bibfield  {author} {\bibinfo {author} {\bibfnamefont {K.}~\bibnamefont
  {Tanaka}}\ and\ \bibinfo {author} {\bibfnamefont {F.}~\bibnamefont
  {Marsiglio}},\ }\bibfield  {title} {\enquote {\bibinfo {title} {{Even-odd and
  super-even effects in the attractive Hubbard model}},}\ }\href {\doibase
  10.1103/PhysRevB.60.3508} {\bibfield  {journal} {\bibinfo  {journal} {Phys.
  Rev. B}\ }\textbf {\bibinfo {volume} {60}},\ \bibinfo {pages} {3508}
  (\bibinfo {year} {1999})}\BibitemShut {NoStop}%
\bibitem [{\citenamefont {Salwen}\ \emph {et~al.}(2004)\citenamefont {Salwen},
  \citenamefont {Sheets},\ and\ \citenamefont {Cotanch}}]{Salwen_aug2004}%
  \BibitemOpen
  \bibfield  {author} {\bibinfo {author} {\bibfnamefont {N.}~\bibnamefont
  {Salwen}}, \bibinfo {author} {\bibfnamefont {S.~A.}\ \bibnamefont {Sheets}},
  \ and\ \bibinfo {author} {\bibfnamefont {S.~R.}\ \bibnamefont {Cotanch}},\
  }\bibfield  {title} {\enquote {\bibinfo {title} {Bcs and attractive hubbard
  model comparative study},}\ }\href {\doibase 10.1103/PhysRevB.70.064511}
  {\bibfield  {journal} {\bibinfo  {journal} {Phys. Rev. B}\ }\textbf {\bibinfo
  {volume} {70}},\ \bibinfo {pages} {064511} (\bibinfo {year}
  {2004})}\BibitemShut {NoStop}%
\bibitem [{\citenamefont {Hu}\ \emph {et~al.}(2010)\citenamefont {Hu},
  \citenamefont {Wang}, \citenamefont {Xianlong}, \citenamefont {Okumura},
  \citenamefont {Igarashi}, \citenamefont {Yamada},\ and\ \citenamefont
  {Machida}}]{Hu_jul2010}%
  \BibitemOpen
  \bibfield  {author} {\bibinfo {author} {\bibfnamefont {J.-H.}\ \bibnamefont
  {Hu}}, \bibinfo {author} {\bibfnamefont {J.-J.}\ \bibnamefont {Wang}},
  \bibinfo {author} {\bibfnamefont {G.}~\bibnamefont {Xianlong}}, \bibinfo
  {author} {\bibfnamefont {M.}~\bibnamefont {Okumura}}, \bibinfo {author}
  {\bibfnamefont {R.}~\bibnamefont {Igarashi}}, \bibinfo {author}
  {\bibfnamefont {S.}~\bibnamefont {Yamada}}, \ and\ \bibinfo {author}
  {\bibfnamefont {M.}~\bibnamefont {Machida}},\ }\bibfield  {title} {\enquote
  {\bibinfo {title} {{Ground-state properties of the one-dimensional attractive
  Hubbard model with confinement: A comparative study}},}\ }\href {\doibase
  10.1103/PhysRevB.82.014202} {\bibfield  {journal} {\bibinfo  {journal} {Phys.
  Rev. B}\ }\textbf {\bibinfo {volume} {82}},\ \bibinfo {pages} {014202}
  (\bibinfo {year} {2010})}\BibitemShut {NoStop}%
\bibitem [{\citenamefont {Schilling}(2018)}]{Schilling_dec2018}%
  \BibitemOpen
  \bibfield  {author} {\bibinfo {author} {\bibfnamefont {C.}~\bibnamefont
  {Schilling}},\ }\bibfield  {title} {\enquote {\bibinfo {title}
  {Communication: {{Relating}} the pure and ensemble density matrix
  functional},}\ }\href {\doibase 10.1063/1.5080088} {\bibfield  {journal}
  {\bibinfo  {journal} {J. Chem. Phys.}\ }\textbf {\bibinfo {volume} {149}},\
  \bibinfo {pages} {231102} (\bibinfo {year} {2018})}\BibitemShut {NoStop}%
\bibitem [{\citenamefont {Schilling}\ and\ \citenamefont
  {Schilling}(2019)}]{Schilling_jan2019}%
  \BibitemOpen
  \bibfield  {author} {\bibinfo {author} {\bibfnamefont {C.}~\bibnamefont
  {Schilling}}\ and\ \bibinfo {author} {\bibfnamefont {R.}~\bibnamefont
  {Schilling}},\ }\bibfield  {title} {\enquote {\bibinfo {title} {Diverging
  {{Exchange Force}} and {{Form}} of the {{Exact Density Matrix
  Functional}}},}\ }\href {\doibase 10.1103/PhysRevLett.122.013001} {\bibfield
  {journal} {\bibinfo  {journal} {Phys. Rev. Lett.}\ }\textbf {\bibinfo
  {volume} {122}},\ \bibinfo {pages} {013001} (\bibinfo {year}
  {2019})}\BibitemShut {NoStop}%
\bibitem [{\citenamefont {Schmidt}\ \emph {et~al.}(2019)\citenamefont
  {Schmidt}, \citenamefont {Benavides-Riveros},\ and\ \citenamefont
  {Marques}}]{Schmidt_jun2019}%
  \BibitemOpen
  \bibfield  {author} {\bibinfo {author} {\bibfnamefont {J.}~\bibnamefont
  {Schmidt}}, \bibinfo {author} {\bibfnamefont {C.~L.}\ \bibnamefont
  {Benavides-Riveros}}, \ and\ \bibinfo {author} {\bibfnamefont {M.~A.~L.}\
  \bibnamefont {Marques}},\ }\bibfield  {title} {\enquote {\bibinfo {title}
  {Reduced density matrix functional theory for superconductors},}\ }\href
  {\doibase 10.1103/PhysRevB.99.224502} {\bibfield  {journal} {\bibinfo
  {journal} {Phys. Rev. B}\ }\textbf {\bibinfo {volume} {99}},\ \bibinfo
  {pages} {224502} (\bibinfo {year} {2019})}\BibitemShut {NoStop}%
\bibitem [{\citenamefont {{Benavides-Riveros}}\ \emph
  {et~al.}(2020)\citenamefont {{Benavides-Riveros}}, \citenamefont {Wolff},
  \citenamefont {Marques},\ and\ \citenamefont
  {Schilling}}]{Benavides-Riveros_may2020}%
  \BibitemOpen
  \bibfield  {author} {\bibinfo {author} {\bibfnamefont {C.~L.}\ \bibnamefont
  {{Benavides-Riveros}}}, \bibinfo {author} {\bibfnamefont {J.}~\bibnamefont
  {Wolff}}, \bibinfo {author} {\bibfnamefont {M.~A.~L.}\ \bibnamefont
  {Marques}}, \ and\ \bibinfo {author} {\bibfnamefont {C.}~\bibnamefont
  {Schilling}},\ }\bibfield  {title} {\enquote {\bibinfo {title} {Reduced
  {{Density Matrix Functional Theory}} for {{Bosons}}},}\ }\href {\doibase
  10.1103/PhysRevLett.124.180603} {\bibfield  {journal} {\bibinfo  {journal}
  {Phys. Rev. Lett.}\ }\textbf {\bibinfo {volume} {124}},\ \bibinfo {pages}
  {180603} (\bibinfo {year} {2020})}\BibitemShut {NoStop}%
\bibitem [{\citenamefont {Schmidt}\ \emph {et~al.}(2021)\citenamefont
  {Schmidt}, \citenamefont {Fadel},\ and\ \citenamefont
  {Benavides-Riveros}}]{Schmidt_sep2021}%
  \BibitemOpen
  \bibfield  {author} {\bibinfo {author} {\bibfnamefont {J.}~\bibnamefont
  {Schmidt}}, \bibinfo {author} {\bibfnamefont {M.}~\bibnamefont {Fadel}}, \
  and\ \bibinfo {author} {\bibfnamefont {C.~L.}\ \bibnamefont
  {Benavides-Riveros}},\ }\bibfield  {title} {\enquote {\bibinfo {title}
  {Machine learning universal bosonic functionals},}\ }\href {\doibase
  10.1103/PhysRevResearch.3.L032063} {\bibfield  {journal} {\bibinfo  {journal}
  {Phys. Rev. Research}\ }\textbf {\bibinfo {volume} {3}},\ \bibinfo {pages}
  {L032063} (\bibinfo {year} {2021})}\BibitemShut {NoStop}%
\bibitem [{\citenamefont {Hohenberg}\ and\ \citenamefont
  {Kohn}(1964)}]{Hohenberg_nov1964}%
  \BibitemOpen
  \bibfield  {author} {\bibinfo {author} {\bibfnamefont {P.}~\bibnamefont
  {Hohenberg}}\ and\ \bibinfo {author} {\bibfnamefont {W.}~\bibnamefont
  {Kohn}},\ }\bibfield  {title} {\enquote {\bibinfo {title} {Inhomogeneous
  {Electron Gas}},}\ }\href {\doibase 10.1103/PhysRev.136.B864} {\bibfield
  {journal} {\bibinfo  {journal} {Phys. Rev.}\ }\textbf {\bibinfo {volume}
  {136}},\ \bibinfo {pages} {B864} (\bibinfo {year} {1964})}\BibitemShut
  {NoStop}%
\bibitem [{\citenamefont {Parr}\ and\ \citenamefont {Yang}(1995)}]{Parr_1995}%
  \BibitemOpen
  \bibfield  {author} {\bibinfo {author} {\bibfnamefont {R.~G.}\ \bibnamefont
  {Parr}}\ and\ \bibinfo {author} {\bibfnamefont {W.}~\bibnamefont {Yang}},\
  }\href@noop {} {\emph {\bibinfo {title} {{Density-Functional Theory of Atoms
  and Molecules}}}}\ (\bibinfo  {publisher} {Oxford University Press},\
  \bibinfo {address} {New York},\ \bibinfo {year} {1995})\BibitemShut {NoStop}%
\bibitem [{\citenamefont {M{\"u}ller}\ \emph {et~al.}(2018)\citenamefont
  {M{\"u}ller}, \citenamefont {T{\"o}ws},\ and\ \citenamefont
  {Pastor}}]{Muller_jul2018}%
  \BibitemOpen
  \bibfield  {author} {\bibinfo {author} {\bibfnamefont {T.~S.}\ \bibnamefont
  {M{\"u}ller}}, \bibinfo {author} {\bibfnamefont {W.}~\bibnamefont
  {T{\"o}ws}}, \ and\ \bibinfo {author} {\bibfnamefont {G.~M.}\ \bibnamefont
  {Pastor}},\ }\bibfield  {title} {\enquote {\bibinfo {title} {Exploiting the
  links between ground-state correlations and independent-fermion entropy in
  the {{Hubbard}} model},}\ }\href {\doibase 10.1103/PhysRevB.98.045135}
  {\bibfield  {journal} {\bibinfo  {journal} {Phys. Rev. B}\ }\textbf {\bibinfo
  {volume} {98}},\ \bibinfo {pages} {045135} (\bibinfo {year}
  {2018})}\BibitemShut {NoStop}%
\bibitem [{\citenamefont {López-Sandoval}\ and\ \citenamefont
  {Pastor}(2000)}]{Lopez-Sandoval_jan2000}%
  \BibitemOpen
  \bibfield  {author} {\bibinfo {author} {\bibfnamefont {R.}~\bibnamefont
  {López-Sandoval}}\ and\ \bibinfo {author} {\bibfnamefont {G.~M.}\
  \bibnamefont {Pastor}},\ }\bibfield  {title} {\enquote {\bibinfo {title}
  {Density-matrix functional theory of the {Hubbard} model: {An} exact
  numerical study},}\ }\href {\doibase 10.1103/PhysRevB.61.1764} {\bibfield
  {journal} {\bibinfo  {journal} {Phys. Rev. B}\ }\textbf {\bibinfo {volume}
  {61}},\ \bibinfo {pages} {1764} (\bibinfo {year} {2000})}\BibitemShut
  {NoStop}%
\bibitem [{\citenamefont {T{\"o}ws}\ \emph {et~al.}(2014)\citenamefont
  {T{\"o}ws}, \citenamefont {Sauban{\`e}re},\ and\ \citenamefont
  {Pastor}}]{Tows2014_tca}%
  \BibitemOpen
  \bibfield  {author} {\bibinfo {author} {\bibfnamefont {W.}~\bibnamefont
  {T{\"o}ws}}, \bibinfo {author} {\bibfnamefont {M.}~\bibnamefont
  {Sauban{\`e}re}}, \ and\ \bibinfo {author} {\bibfnamefont {G.~M.}\
  \bibnamefont {Pastor}},\ }\bibfield  {title} {\enquote {\bibinfo {title}
  {Density-matrix functional theory of strongly correlated fermions on lattice
  models and minimal-basis hamiltonians},}\ }\href {\doibase
  10.1007/s00214-013-1422-0} {\bibfield  {journal} {\bibinfo  {journal} {Theor.
  Chem. Acc.}\ }\textbf {\bibinfo {volume} {133}},\ \bibinfo {pages} {1}
  (\bibinfo {year} {2014})}\BibitemShut {NoStop}%
\bibitem [{\citenamefont {Töws}\ and\ \citenamefont
  {Pastor}(2011)}]{Tows_jun2011}%
  \BibitemOpen
  \bibfield  {author} {\bibinfo {author} {\bibfnamefont {W.}~\bibnamefont
  {Töws}}\ and\ \bibinfo {author} {\bibfnamefont {G.~M.}\ \bibnamefont
  {Pastor}},\ }\bibfield  {title} {\enquote {\bibinfo {title} {Lattice density
  functional theory of the single-impurity {Anderson} model: {Development} and
  applications},}\ }\href {\doibase 10.1103/PhysRevB.83.235101} {\bibfield
  {journal} {\bibinfo  {journal} {Phys. Rev. B}\ }\textbf {\bibinfo {volume}
  {83}},\ \bibinfo {pages} {235101} (\bibinfo {year} {2011})}\BibitemShut
  {NoStop}%
\bibitem [{\citenamefont {Levy}(1982)}]{Levy_sep1982}%
  \BibitemOpen
  \bibfield  {author} {\bibinfo {author} {\bibfnamefont {M.}~\bibnamefont
  {Levy}},\ }\bibfield  {title} {\enquote {\bibinfo {title} {Electron densities
  in search of {Hamiltonians}},}\ }\href {\doibase 10.1103/PhysRevA.26.1200}
  {\bibfield  {journal} {\bibinfo  {journal} {Phys. Rev. A}\ }\textbf {\bibinfo
  {volume} {26}},\ \bibinfo {pages} {1200} (\bibinfo {year}
  {1982})}\BibitemShut {NoStop}%
\bibitem [{\citenamefont {Lieb}(1983)}]{Lieb_sep1983}%
  \BibitemOpen
  \bibfield  {author} {\bibinfo {author} {\bibfnamefont {E.~H.}\ \bibnamefont
  {Lieb}},\ }\bibfield  {title} {\enquote {\bibinfo {title} {{Density
  Functionals for Coulomb Systems}},}\ }\href {\doibase 10.1002/qua.560240302}
  {\bibfield  {journal} {\bibinfo  {journal} {Int. J. Quant. Chem.}\ }\textbf
  {\bibinfo {volume} {24}},\ \bibinfo {pages} {243} (\bibinfo {year}
  {1983})}\BibitemShut {NoStop}%
\bibitem [{\citenamefont {Valone}(1980{\natexlab{a}})}]{Valone_aug1980}%
  \BibitemOpen
  \bibfield  {author} {\bibinfo {author} {\bibfnamefont {S.~M.}\ \bibnamefont
  {Valone}},\ }\bibfield  {title} {\enquote {\bibinfo {title} {Consequences of
  extending $1$-matrix energy functionals from pure-state representable to all
  ensemble representable $1$ matrices},}\ }\href {\doibase 10.1063/1.440249}
  {\bibfield  {journal} {\bibinfo  {journal} {J. Chem. Phys.}\ }\textbf
  {\bibinfo {volume} {73}},\ \bibinfo {pages} {1344} (\bibinfo {year}
  {1980}{\natexlab{a}})}\BibitemShut {NoStop}%
\bibitem [{\citenamefont {Valone}(1980{\natexlab{b}})}]{Valone_nov1980}%
  \BibitemOpen
  \bibfield  {author} {\bibinfo {author} {\bibfnamefont {S.~M.}\ \bibnamefont
  {Valone}},\ }\bibfield  {title} {\enquote {\bibinfo {title} {A one-to-one
  mapping between one-particle densities and some {$n$}-particle ensembles},}\
  }\href {\doibase 10.1063/1.440656} {\bibfield  {journal} {\bibinfo  {journal}
  {J. Chem. Phys.}\ }\textbf {\bibinfo {volume} {73}},\ \bibinfo {pages} {4653}
  (\bibinfo {year} {1980}{\natexlab{b}})}\BibitemShut {NoStop}%
\bibitem [{Note1()}]{Note1}%
  \BibitemOpen
  \bibinfo {note} {The domain of definition of $W[\protect \bm {\gamma }]$
  comprises not only density matrices which can be derived from some
  ground-state of the Hubbard model (i.e., the so-called ground-state
  representable $\protect \bm {\gamma }$) but the much broader set of
  ensemble-representable density matrices, which can be derived from an
  arbitrary mixed state $\protect \mathaccentV {hat}05E{\rho }$. The necessary
  and sufficient condition for a density matrix to be ensemble representable is
  simply that all its eigenvalues $\eta _{\protect \bm {k}\sigma }$ are
  comprised between zero and one.}\BibitemShut {Stop}%
\bibitem [{\citenamefont {Collins}(1993)}]{Collins1993}%
  \BibitemOpen
  \bibfield  {author} {\bibinfo {author} {\bibfnamefont {D.~M.}\ \bibnamefont
  {Collins}},\ }\bibfield  {title} {\enquote {\bibinfo {title} {Entropy
  maximizations on electron density},}\ }\href {\doibase
  10.1515/zna-1993-1-218} {\bibfield  {journal} {\bibinfo  {journal} {Z.
  Naturforsch. A}\ }\textbf {\bibinfo {volume} {48}},\ \bibinfo {pages} {68}
  (\bibinfo {year} {1993})}\BibitemShut {NoStop}%
\bibitem [{Note2()}]{Note2}%
  \BibitemOpen
  \bibinfo {note} {In Ref.~\cite {Collins1993} the importance of the growing
  statistical uncertainty in the occupation-number distribution resulting from
  increasing correlations has been clearly identified. The arguments presented
  here in order to derive Eq.~(\ref {eq:9}) in the weakly correlated limit are
  very similar to those formulated in Collins' work. However, Eq.~(\ref {eq:9})
  is profoundly different physically from the information entropy $S = -\DOTSB
  \sum@ \slimits@ _k \eta _k \protect \qopname \relax o{ln}(\eta _k)$
  considered in Ref.~\cite {Collins1993}. Eq.~(\ref {eq:9}) represents the
  entropy of a many-body system of non-interacting fermions having the same
  occupation-number distribution $\eta _{k\sigma }$ as the interacting system
  under consideration. Thus, the fermionic character of the particles is
  correctly taken into account and the fundamental electron-hole symmetry of
  $W[\protect \bm {\eta }]$ is respected, which does not hold for Collins'
  conjecture.}\BibitemShut {Stop}%
\bibitem [{\citenamefont {Flores-Gallegos}(2016)}]{Flores2016}%
  \BibitemOpen
  \bibfield  {author} {\bibinfo {author} {\bibfnamefont {N.}~\bibnamefont
  {Flores-Gallegos}},\ }\bibfield  {title} {\enquote {\bibinfo {title}
  {Informational energy as a measure of electron correlation},}\ }\href
  {\doibase 10.1016/j.cplett.2016.10.075} {\bibfield  {journal} {\bibinfo
  {journal} {Chem. Phys. Lett.}\ }\textbf {\bibinfo {volume} {666}},\ \bibinfo
  {pages} {62} (\bibinfo {year} {2016})}\BibitemShut {NoStop}%
\bibitem [{\citenamefont {Wang}\ \emph {et~al.}(2021)\citenamefont {Wang},
  \citenamefont {Knowles},\ and\ \citenamefont {Wang}}]{Wang2021}%
  \BibitemOpen
  \bibfield  {author} {\bibinfo {author} {\bibfnamefont {Y.}~\bibnamefont
  {Wang}}, \bibinfo {author} {\bibfnamefont {P.~J.}\ \bibnamefont {Knowles}}, \
  and\ \bibinfo {author} {\bibfnamefont {J.}~\bibnamefont {Wang}},\ }\bibfield
  {title} {\enquote {\bibinfo {title} {Information entropy as a measure of the
  correlation energy associated with the cumulant},}\ }\href {\doibase
  10.1103/PhysRevA.103.062808} {\bibfield  {journal} {\bibinfo  {journal}
  {Phys. Rev. A}\ }\textbf {\bibinfo {volume} {103}},\ \bibinfo {pages}
  {062808} (\bibinfo {year} {2021})}\BibitemShut {NoStop}%
\bibitem [{Note3()}]{Note3}%
  \BibitemOpen
  \bibinfo {note} {The electron-hole transformation relating positive-$U$ and
  negative-$U$ Hubbard models is obtained by introducing the following
  hole-annihilation operators: $\protect \ca [h]{i\protect \ensuremath
  {\delimiter "3222378 }}=\protect \ca {i\protect \ensuremath {\delimiter
  "3222378 }}$ and $\protect \ca [h]{i\protect \ensuremath {\delimiter "3223379
  }}=\protect \ct {i\protect \ensuremath {\delimiter "3223379 }}$. It implies a
  change of sign in the down-spin hopping integrals and therefore leaves
  $\protect \mathaccentV {hat}05E{H}$ invariant only for bipartite lattices, in
  which case one sets $\protect \ca [h]{i\protect \ensuremath {\delimiter
  "3223379 }}=-\protect \ct {i\protect \ensuremath {\delimiter "3223379 }}$ for
  $i$ belonging to one of the two sublattices. In addition an irrelevant shift
  of the total energy is obtained. Notice that if $\Delta S_z = N_{\protect
  \ensuremath {\delimiter "3223379 }} - N_a/2\not =0$ a change in the total
  spin-polarization $S_z^{(h)} = S_z+\Delta S_z$ is involved.}\BibitemShut
  {Stop}%
\end{thebibliography}%

\end{document}